\documentclass[12pt]{article}%
\usepackage{amsmath}
\usepackage{amsfonts}
\usepackage{amssymb}
\usepackage{graphicx}
\usepackage{subfigure}%
\newtheorem{theorem}{Theorem}

\newtheorem{definition}[theorem]{Definition}

\newtheorem{lemma}[theorem]{Lemma}

\newenvironment{proof}[1][Proof]{\noindent\textbf{#1.} }{\ \rule{0.5em}{0.5em}}
\begin{document}

\title{Is Symplectic-Energy-Momentum Integration Well-Posed?\thanks{Dedicated to the
memory of my father Shibberu Wolde Mariam.}}
\author{Yosi Shibberu\\Mathematics Department\\Rose-Hulman Institute of Technology\\Terre Haute, IN 47803\\shibberu@rose-hulman.edu\\www.rose-hulman.edu/$\thicksim$shibberu/DTH\_Dynamics/DTH\_Dynamics.htm}
\maketitle

\begin{abstract}
We provide new existence and uniqueness results for the discrete-time Hamilton
(DTH) equations of a symplectic-energy-momentum (SEM) integrator. In
particular, we identify points in extended-phase space where the DTH equations
of SEM integration have no solution for arbitrarily small time steps. We use
the nonlinear pendulum to illustrate the main ideas.

\begin{description}
\item[Key Words] DTH dynamics, symplectic energy momentum integrator,
variational integrator, discrete mechanics, discrete time Hamiltonian,
discrete variational principles, principle of least action, energy conserving
methods, extended phase space, midpoint method, variable-time step, adaptive.

\end{description}
\end{abstract}

\section{Background}

Is symplectic-energy-momentum integration well-posed? Loosely speaking, the
answer is no. Points exist in the extended phase-space of a Hamiltonian system
where the equations of a symplectic-energy-momentum (SEM) integrator have no
solution for arbitrarily small time steps. Before considering this question in
more detail, we provide a brief review of SEM integration.

Hamiltonian dynamics is at the heart of modern physics and arises naturally in
applications such as optimal control theory and geometric optics. Hamiltonian
dynamics is also the inspiration for the relatively new field of symplectic
geometry. A symplectic-energy-momentum (SEM) integrator is a numerical
integrator that preserves the following key properties associated with
Hamiltonian dynamics: i) The integrator is symplectic. ii) The integrator
exactly conserves energy (the Hamiltonian function). iii) The integrator
exactly preserves \textquotedblleft linear\textquotedblright\ symmetries (e.g.
linear and angular momentum in Cartesian coordinates). The term
\textquotedblleft symplectic-energy-momentum integrator\textquotedblright\ was
coined and popularized by Kane, Marsden and Ortiz \cite{Kane-99}. See also
Chen, Guo and Wu \cite{Chen-03} for related work on higher-order,
symplectic-energy integrators. Guibout and Bloch \cite{Guibout-04} have
developed a general framework for deriving many of the published symplectic
integrators, including SEM integrators.

The author's work on SEM integration---known as discrete-time, Hamiltonian
(DTH) dynamics---predates the work of Kane, et al. \cite{Kane-99}. DTH
dynamics originated from an effort to obtain the exact energy and momentum
conserving properties of the discrete mechanics of Greenspan
\cite{Greenspan-74}, \cite{Greenspan-80}, from the variational principle used
in the discrete mechanics of Lee \cite{Lee-83}, \cite{Lee-87}. DTH dynamics
was proved in 1994 (see Shibberu \cite{Shibberu-94}, \cite{Shibberu-98}) to be
symplectic and hence a SEM integrator.

In the extended-phase space formulation of Hamiltonian dynamics, time is
treated as a generalized coordinate on equal footing with the position
coordinates. The momentum conjugate to time is introduced as an additional
generalized coordinate. The principle of least (stationary) action takes a
particularly simple form in extended-phase space. But, despite its aesthetic
appeal, the extended-phase space formulation of the principle of least action
is not widely used because it leads to indeterminate equations of motion
\cite{Lanczos-70}, \cite{Goldstein-80}, \cite{Shibberu-98}. Lee \cite{Lee-83},
\cite{Lee-87}, described a discretization of Lagrangian dynamics that appeared
to remove this indeterminacy. D'Innocenzo, Renna and Rotelli
\cite{DInnocenzo-87} modified Lee's discretization and achieved exact energy
conservation. SEM integration is based on a related, but more general,
discretization developed independently of D'Innocenzo et al.
\cite{DInnocenzo-87} in Shibberu \cite{Shibberu-92}.

An important theorem due to Ge (see \cite{Ge-91} and citation in \cite{Ge-88})
illustrates the difficulty of formulating a symplectic integrator which
exactly conserves energy. Roughly speaking, Ge's Theorem says that a general,
energy conserving, symplectic discretization of Hamiltonian dynamics, must
reproduce a reparametrization of the exact dynamics. Why SEM integration does
not violate Ge's Theorem was explained for the first time in Shibberu
\cite{Shibberu-97}.\footnote{The explanation given in Kane et al.
\cite{Kane-99} of why symplectic-energy-momentum integration does not violate
Ge's Theorem is incorrect. A variable-time step symplectic integrator can be
reformulated in extended-phase space as a constant-time step symplectic
integrator. Therefore, Ge's Theorem holds true even for variable time-step
symplectic integrators. See the discussion of
\mbox{Hairer's \cite{Hairer-97}}
\textquotedblleft meta-algorithm\textquotedblright\ for variable time-step
symplectic integrators given in the last section of Shibberu
\cite{Shibberu-98}.}

This article is concerned with the following question. Under what conditions
are the DTH equations of SEM integration well-posed? We will prove results
which generalize the existence and uniqueness results first proved in Shibberu
\cite{Shibberu-92}. The existence and uniqueness results in this article are
for nonlinear Hamiltonian systems and are local in nature. A global result for
linear Hamiltonian systems was proved in Shibberu \cite{Shibberu-92},
\cite{Shibberu-98}.

\section{Example: The Nonlinear Pendulum}

In this section, we illustrate the main ideas of this article using the
nonlinear pendulum as an example. We begin by describing how the DTH equations
of Hamiltonian dynamics are derived. Then we consider the existence and
uniqueness of solutions to the DTH equations.

Let $z=(q,p)^{\top}$ where $q=\left(  q_{1},\ldots,q_{n},t\right)  ^{\top}$
and $p=\left(  p_{1},\ldots,p_{n},\wp\right)  ^{\top}$ are the extended phase
space, position and momentum coordinates of an $n$ degree-of-freedom
Hamiltonian dynamical system with Hamiltonian function $\mathcal{H}(z).$ The
position coordinate $t$ represents time and the momentum coordinate $\wp$
represents the momentum conjugate to time. (See \cite{Lanczos-70},
\cite{Goldstein-80} or \cite{Shibberu-94} for a detailed description of $\wp
.$) We represent the motion of a discrete-time Hamiltonian dynamical system by
a piecewise-linear, continuous trajectory in extended-phase space where
$z_{k},$ $k=0,\ldots,N$ are the vertices of the trajectory and $\overline
{z}_{k},$ $k=0,\ldots,N-1$ are the midpoints of the linear segments of the trajectory.

Define the \emph{one-step action} of a discrete-time Hamiltonian dynamical
system to be the function $\mathcal{A}(z_{k},z_{k+1})=\frac{1}{2}\Delta
q_{k}{}^{\top}\Delta p_{k}.$ (The motivation for choosing this definition for
the discrete action is given in Shibberu \cite{Shibberu-05}.) The dynamics of
a discrete-time Hamiltonian dynamical system is determined by the following
variational principle.

\begin{definition}
[DTH Principle of Stationary Action]\label{DTH_principle}The one-step action
$\mathcal{A}(z_{k},z_{k+1}),$ $k=0,1,\ldots N-1$, is stationary along a DTH
trajectory for variations which fix $q_{k}\ $and $p_{k+1}$ and satisfy the
Hamiltonian constraint $\mathcal{H}(\overline{z}_{k})=0.$
\end{definition}

The DTH equations of SEM integration are determined by Definition
\ref{DTH_principle}.

\begin{theorem}
[DTH Equations]\label{DTH_equations}A DTH trajectory is determined by the
following equations:%
\begin{subequations}
\begin{align}
\Delta z_{k}  &  =\lambda_{k}J\mathcal{H}_{z}(\overline{z}_{k}%
)\ \label{DTH_equations_1}\\
\mathcal{H}(\overline{z}_{k})  &  =0 \label{DTH_equations_2}%
\end{align}
where $J=\left(
\begin{array}
[c]{rr}%
0 & I\\
-I & 0
\end{array}
\right)  $and $I$ is the $n+1$ dimensional identity matrix.
\end{subequations}
\end{theorem}

Theorem \ref{DTH_equations} is proved in Shibberu \cite{Shibberu-05}. See also
Shibberu \cite{Shibberu-98} for the proof that the DTH equations
(\ref{DTH_equations_1})--(\ref{DTH_equations_2}) preserve
symplectic-energy-momentum properties and are coordinate invariant under
linear symplectic coordinate transformations.

For sufficiently small time steps, a sufficient condition for the existence
and (local) uniqueness of solutions to equations (\ref{DTH_equations_1}%
)--(\ref{DTH_equations_2}) is the condition $\psi(z_{k})\neq0$ where
$\psi=(J\mathcal{H}_{z})^{\top}\mathcal{H}_{zz}(J\mathcal{H}_{z})$ Shibberu
\cite{Shibberu-92}. The new existence and uniqueness results proved in this
article include points where $\psi(z_{k})$ may equal zero, but the Poisson
bracket $\left.  [\psi,\mathcal{H]}\right\vert _{z_{k}}$ is not equal to zero.
Smoothness requirements on the Hamiltonian function are also weakened from
$\mathcal{H\in}C^{3}(U)$ to $\mathcal{H\in}C^{2}(U)$ where $U\subset\Re
^{2n+2}$ is on open set in extended-phase space. (See Theorem
\ref{main_result} on page \pageref{main_result} for the main result of this article).

Consider now a nonlinear pendulum with extended-phase space Hamiltonian
function $\mathcal{H}(q,p,\wp)=\wp+\frac{1}{2}p-\cos(q).$ (Recall that $\wp$
is the momentum conjugate to time.) The corresponding discrete-time Hamilton
(DTH) equations are%
\begin{subequations}
\begin{align*}
\Delta q_{k}  &  =\lambda_{k}\overline{p}_{k}\\
\Delta t_{k}  &  =\lambda_{k}\\
\Delta p_{k}  &  =-\lambda_{k}\sin(\overline{q}_{k})\\
\Delta\wp_{k}  &  =0\\
\overline{\wp}_{k}+\frac{1}{2}\overline{p}_{k}^{2}-\cos(\overline{q}_{k})  &
=0.
\end{align*}
Figure \ref{phase_portrait} is a plot of a DTH trajectory determined by the
above equations and projected onto the phase portrait of the pendulum. Observe
that the linear segments of the DTH trajectory are tangent to an energy
conserving manifold of the pendulum. (We stress that the size of the initial
time step, $\lambda_{0},\ $is determined by the initial condition
$z_{0}=(q_{0},t_{0},p_{0,}\wp_{0}).$) The v-shaped curves in Figure
\ref{phase_portrait} are points where $\psi(z)$ equals zero. The horizontal
and vertical lines are points where the Poisson bracket $[\psi,\mathcal{H]}%
\ $equals zero. From Figure \ref{phase_portrait}, we see that the existence
and uniqueness results in this article apply to all the points in phase space
except the equilibrium points where both $\psi(z)$ and $[\psi,\mathcal{H]}$
are equal to zero.
\begin{figure}
[ptb]
\begin{center}
\includegraphics[
height=2.2857in,
width=3.0286in
]%
{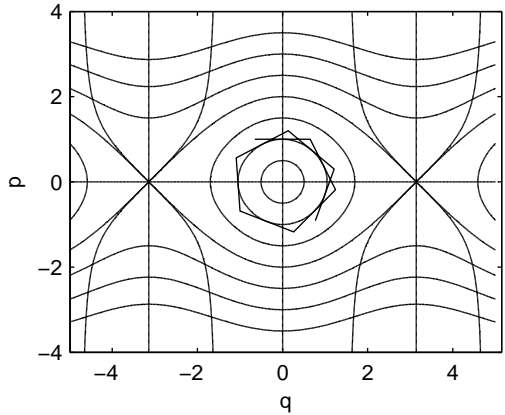}%
\caption{A DTH trajectory of a nonlinear pendulum. The v-shaped curves
correspond to points where $\psi(z)=0$ and the horizontal and vertical lines
correspond to points where $[\psi,\mathcal{H]}=0.$}%
\label{phase_portrait}%
\end{center}
\end{figure}

Let $\psi_{k}=$ $\psi(z_{k}).$ We will show that, for points where $\psi
_{k}\neq0,$ the magnitude and sign of $\mathcal{H}_{k}/\psi_{k}$ is key to
determining if a solution to the DTH equations exists and is locally unique.
In particular, if $\mathcal{H}_{k}/\psi_{k}<0,$ and $\psi_{k}$ is sufficiently
large, then no solution exists. If $\psi_{k}=0,$ the quantity $\mathcal{H}%
_{k}/\psi_{k}^{\prime},\ $where $\psi^{\prime}=[\psi,\mathcal{H]},$ plays a
similar role in determining existence and uniqueness. In the neighborhood of
points where $\psi$ changes sign, a DTH trajectory bifurcates giving rise to
\textquotedblleft ghost trajectories\textquotedblright. Ghost trajectories are
discussed in more detail in section \ref{ghost_trj}.

The outline of this article is as follows. In section
\ref{decoupling_function}, we use the Newton-Kantorovich Theorem to prove the
existence and uniqueness of a function $\overline{z}(\lambda,z_{k})$
implicitly defined by equation (\ref{DTH_equations_1}). We use the function
$\overline{z}(\lambda,z_{k})$ to decouple equation (\ref{DTH_equations_2})
from equation (\ref{DTH_equations_1}). In section \ref{cubic_bounds}, we
derive a cubic approximation of the Hamiltonian constraint function
$g(\lambda,z_{k})=\mathcal{H(}\overline{z}(\lambda,z_{k})\mathcal{)}$. In
section \ref{multipliers}, we identify intervals where $g(\lambda,z_{k})$ is
monotonic increasing/decreasing with respect to $\lambda$. Using monotonicity
and the Intermediate Value Theorem, we prove the existence and uniqueness of
Lagrange multipliers satisfying the decoupled, Hamiltonian constraint equation
$g(\lambda,z_{k})=0$. The existence and uniqueness results for Lagrange
multipliers is used in section \ref{EU_section} to prove the existence and
uniqueness of DTH trajectories. SEM integration is shown, under certain
conditions, to be well-posed. Finally, in section \ref{ghost_trj}, we discuss
ghost trajectories and the need to regularize the DTH equations of SEM integration.

\section{Existence of a Decoupling Function}

Consider the DTH equations (\ref{DTH_equations_1})--(\ref{DTH_equations_2}).
Equation (\ref{DTH_equations_1}) can be rewritten as $f(\lambda,z_{k}%
,\overline{z}_{k})=\overline{z}_{k}-z_{k}-\frac{1}{2}\lambda J\mathcal{H}%
_{z}\left(  \overline{z}\right)  =0$ where $\overline{z}_{k}=\frac{1}%
{2}(z_{k+1}+z_{k}).$ In Theorem \ref{Decoupling} below, we prove that if the
Hamiltonian function $\mathcal{H}(z)$ satisfies certain conditions, then there
exists a smooth function $\overline{z}(\lambda,z_{k})$ such that
$f(\lambda,z_{k},\overline{z}(\lambda,z_{k}))=0$ for all $\lambda\in
\lbrack-\lambda_{\delta},\lambda_{\delta}]$ and $z_{k}\in U_{\delta}$ where
$\lambda_{\delta}$ and $U_{\delta}$ are specified in Theorem \ref{Decoupling}.
The function $\overline{z}(\lambda,z_{k})$ is used in section
\ref{multipliers} to decouple equation (\ref{DTH_equations_2}) from equation
(\ref{DTH_equations_1}). We begin by stating two standard results in numerical
analysis, the Newton-Kantorovich Theorem \cite{Ortega-72} and the Matrix
Perturbation Lemma \cite{Golub-89}.
\end{subequations}
\begin{theorem}
[Newton-Kantorovich Theorem]\label{NK_Thm}Consider the function
$f:U\rightarrow\mathcal{R}^{n}$ where $U\subset\mathcal{R}^{n}$ is open.
Assume $f\in C^{1}(U)$ and $\left\Vert f_{x}(x_{2})-f_{x}(x_{1})\right\Vert
\leq\gamma\left\Vert x_{2}-x_{1}\right\Vert $ for all $x_{1},x_{2}\in U.$
Assume there exists a point $x_{0}\in U$ and constants $\beta>0,$ $\eta>0$
such that $\left\Vert f_{x}(x_{0})^{-1}\right\Vert \leq\beta$ and $\left\Vert
f_{x}(x_{0})^{-1}f(x_{0})\right\Vert \leq\eta$. Assume $\alpha<\frac{1}{2}$
where $\alpha=\beta\gamma\eta.$ Define $r_{-}=\left(  1-\sqrt{1-2\alpha
}\right)  /\beta\gamma$ and $r_{+}=\left(  1+\sqrt{1-2\alpha}\right)
/\beta\gamma.$ If the close ball $\overline{B}(x_{0},r_{-})$ $\subset U,$ then
the Newton iterates $x^{(i)},$ defined by $x^{(i+1)}=x^{(i)}-f_{x}%
(x^{(i)})^{-1}f(x^{(i)}),$ $i=0,1,\ldots,$ with $x^{(0)}=x_{0},$ are well
defined and converge to $x_{\ast}\in\overline{B}(x_{0},r_{-})$ where $x_{\ast
}$ is the unique solution of $f(x)=0$ in $\overline{B}(x_{0},r_{+})\cap U.$
\end{theorem}

\begin{lemma}
[Matrix Perturbation Lemma]\label{Pert_Lemma}Assume the identity matrix $I$ is
perturbed by the matrix $E.$ If $\left\Vert E\right\Vert \nolinebreak%
<\nolinebreak1,$ then $\left(  I-E\right)  ^{-1}$ exists, $\left(  I-E\right)
^{-1}=\sum_{n=0}^{\infty}E^{n}$ and $\left\Vert \left(  I-E\right)
^{-1}\right\Vert \nolinebreak<1\nolinebreak/\left(  1-\left\Vert E\right\Vert
\right)  .$
\end{lemma}

\begin{theorem}
[Decoupling Function]\label{Decoupling}Consider the extended-phase space
Hamiltonian function $\mathcal{H}\in C^{2}(U)$ where $U\subset\mathcal{R}%
^{2n+2}$ is open. Assume $\left\Vert \mathcal{H}_{z}(z)\right\Vert \leq M_{1}$
and $\left\Vert \mathcal{H}_{zz}(z)\right\Vert \leq M_{2}$ for all $z\in U.$
Assume $\left\Vert \mathcal{H}_{zz}(z_{1})-\mathcal{H}_{zz}(z_{2})\right\Vert
\leq\gamma_{H}\left\Vert z_{1}-z_{2}\right\Vert $ for all $z_{1,}z_{2}\in U.$
Let $\mathcal{\lambda}_{\delta}=\min(1/M_{2},1/\gamma_{H},\left(
1-(1-\delta)^{2}\right)  /2M_{1})$ and $U_{\delta}=\left\{  z:\overline
{B}(z,\delta)\subset U\right\}  .$ Define $f(\lambda,z,\overline{z}%
)=\overline{z}-z-\frac{1}{2}\lambda J\mathcal{H}_{z}\left(  \overline
{z}\right)  .$ Then, there exists a $\delta,$ where $0<\delta<1,$ and there
exist a continuously differentiable function $\overline{z}:[-\mathcal{\lambda
}_{\delta},\mathcal{\lambda}_{\delta}]\times U_{\delta}\rightarrow
\mathcal{R}^{2n+2},$ such that $f(\lambda,z,\overline{z}(\lambda,z))=0\ $for
all $(\lambda,z)\in\lbrack-\mathcal{\lambda}_{\delta},\mathcal{\lambda
}_{\delta}]\times U_{\delta}.$
\end{theorem}

\begin{proof}
First we show that for $\left\vert \lambda\right\vert \leq1/M_{2},\ $
$f_{\overline{z}}^{-1}$ exists and is bounded. Since $f_{\overline{z}}=I-E,$
where $E=$ $\frac{1}{2}\lambda J\mathcal{H}_{zz}$, and since $\left\Vert
E\right\Vert \leq\frac{1}{2}\left(  1/M_{2}\right)  M_{2}=\frac{1}{2}<1,$ by
the Matrix Perturbation Lemma, $f_{\overline{z}}^{-1}$ exists and $\left\Vert
f_{\overline{z}}^{-1}\right\Vert <\beta,$ where $\beta=1/\left(  1-\frac{1}%
{2}\right)  =2.$ Next, we show that for $\left\vert \lambda\right\vert
\leq1/\gamma_{H}$, $f_{\overline{z}}$ is Lipschitz with respect to
$\overline{z}$ with Lipschitz constant $\gamma=\frac{1}{2}$. We have
\begin{align*}
\left\Vert f_{\overline{z}}(\lambda,\overline{z}_{2},z)-f_{\overline{z}%
}(\lambda,\overline{z}_{1},z)\right\Vert  &  =\left\Vert \frac{1}{2}\lambda
J\mathcal{H}_{zz}(\overline{z}_{2})-\frac{1}{2}\lambda J\mathcal{H}%
_{zz}(\overline{z}_{1})\right\Vert \\
&  \leq\frac{1}{2}\frac{1}{\gamma_{H}}\gamma_{H}\left\Vert \overline{z}%
_{2}-\overline{z}_{1}\right\Vert \\
&  =\frac{1}{2}\left\Vert \overline{z}_{2}-\overline{z}_{1}\right\Vert .
\end{align*}
Now consider using Newton's iteration to solve $f(\lambda,z_{k},\overline
{z})=0$ for $\overline{z}.$ If we initialize the iteration with $\overline
{z}^{(0)}=z_{k},$ we have $\eta=\left\Vert f_{\overline{z}}^{-1}(\lambda
,z_{k},z_{k})f(\lambda,z_{k},z_{k})\right\Vert \leq\left\Vert f_{\overline{z}%
}^{-1}\right\Vert \left\Vert \frac{1}{2}\lambda J\mathcal{H}_{z}\right\Vert .$
Let $\mathcal{\lambda}_{\delta}=\min(1/M_{2},1/\gamma_{H},\left(
1-(1-\delta)^{2}\right)  /2M_{1}).$ For $\left\vert \lambda\right\vert
\leq\mathcal{\lambda}_{\delta},$ it follows that $\eta<2\frac{1}{2}\left(
1-(1-\delta)^{2}\right)  /\left(  2M_{1}\right)  M_{1}=\left(  1-(1-\delta
)^{2}\right)  /2.$ For $0<\delta<1,$ we have then that $\ \alpha=\beta
\gamma\eta<2\frac{1}{2}\left(  1-(1-\delta)^{2}\right)  /2<\frac{1}{2}.$
Therefore,%
\[
r_{-}=\frac{1-\sqrt{1-2\alpha}}{\beta\gamma}<\frac{1-\sqrt{1-2\left(
\frac{1-(1-\delta)^{2}}{2}\right)  }}{2\frac{1}{2}}=\delta.
\]
It follows that for $z_{k}\in U_{\delta},$ $\overline{B}(z_{k},r_{-}%
)\subset\overline{B}(z_{k},\delta)\subset U.$ By the Newton-Kantorovich
Theorem, the function $\overline{z}(\lambda,z_{k})$ is well defined on
$[-\mathcal{\lambda}_{\delta},\mathcal{\lambda}_{\delta}]\times U_{\delta}$
and $f(\lambda,z_{k},\overline{z}(\lambda,z_{k}))\equiv0.$ (We assume $\delta$
is chosen small enough that $U_{\delta}$ is nonempty.) The Implicit Function
Theorem implies $\overline{z}(\lambda,z_{k})$ is continuously differentiable.
\end{proof}

\section{Cubic Approximation of the Hamiltonian Constraint
\label{cubic_bounds}}

Given $\lambda_{k}\in\lbrack-\mathcal{\lambda}_{\delta},\mathcal{\lambda
}_{\delta}]$ and $z_{k}\in U_{\delta},$ Theorem \ref{Decoupling} implies there
exists a point $\overline{z}_{k}=$ $\overline{z}(\lambda_{k},z_{k})$ and a
point $z_{k+1}=2\overline{z}_{k}-$ $z_{k},\ $such that $\lambda_{k},$ $z_{k}$
and $z_{k+1}$ satisfy the first DTH equation, $\Delta z_{k}=\lambda
_{k}J\mathcal{H}_{z}(\overline{z}_{k})\ .$ We use $\overline{z}(\lambda
,z_{k})$ to decouple the second DTH equation, $\mathcal{H}(\overline{z}%
_{k})=0,$ from the first DTH equation by defining the function $g(\lambda
,z_{k})=\mathcal{H}(\overline{z}(\lambda,z_{k}))$ and replacing the second
equation with the equation $g(\lambda,z_{k})=0.$

In this section, we determine a cubic approximation of $g(\lambda,z_{k})$ as a
function of $\lambda.$ Obtaining this approximation is made difficult by the
fact that the function $\overline{z}(\lambda,z_{k})$ is only implicitly
defined. We will see that the linear term in the cubic approximation of
$g(\lambda,z_{k})$ is always equal to zero. The analysis of DTH dynamics is
also complicated by this fact since we are forced to consider the effects of
the quadratic and even cubic term in the cubic approximation of $g(\lambda
,z_{k}).$

The outline of this section is as follows. In Lemma \ref{z_lambda} below, we
show that $\overline{z}_{\lambda}(\lambda,z_{k})$ is Lipschitz continuous with
respect to $\lambda.$ In Lemma \ref{g_lambda} we define the important function
$\psi(z)=\left(  J\mathcal{H}_{z}\right)  ^{\top}\mathcal{H}_{zz}\left(
J\mathcal{H}_{z}\right)  \ $and we approximate the partial derivative
$\partial g(\lambda,z_{k})/\partial\lambda$ by the simpler function $-\frac
{1}{4}\lambda\,h(\lambda,z_{k})$ where $h(\lambda,z_{k})=$ $\psi
\mathcal{(}\overline{z}(\lambda,z_{k})).$ In Lemma \ref{h_lambda}, we prove
that $\partial h(\lambda,z_{k})/\partial\lambda$ is Lipschitz continuous with
respect to $\lambda.$ Finally, in Lemma \ref{g_taylor}, we determine a cubic
approximate of $g(\lambda,z_{k}).$

\begin{lemma}
\label{z_lambda}For $\lambda_{1},\lambda_{2}\in\lbrack-\mathcal{\lambda
}_{\delta},\mathcal{\lambda}_{\delta}]$ and $z_{k}\in U_{\delta},$
\[
\left\Vert \overline{z}_{\lambda}(\lambda_{2},z_{k})-\overline{z}_{\lambda
}(\lambda_{1},z_{k})\right\Vert \leq\gamma_{z}\left\vert \lambda_{2}%
-\lambda_{1}\right\vert \text{ }%
\]
where $\gamma_{z}=2M_{1}M_{2}+\,M_{1}^{2}.$
\end{lemma}

\noindent The proof is given in the appendix.

\begin{lemma}
\label{g_lambda}Define the functions $g(\lambda,z_{k})=\mathcal{H(}%
\overline{z}(\lambda,z_{k})),$ $\psi(z)=\left(  J\mathcal{H}_{z}\right)
^{\top}\mathcal{H}_{zz}\left(  J\mathcal{H}_{z}\right)  \ $and $h(\lambda
,z_{k})=\psi\mathcal{(}\overline{z}(\lambda,z_{k})).$ Then, for $\left\vert
\lambda\right\vert \leq\lambda_{\delta}\ $and $z_{k}\in U_{\delta},$
\[
\left\vert \frac{\partial g(\lambda,z_{k})}{\partial\lambda}-\left(  -\frac
{1}{4}\lambda h(\lambda,z_{k})\right)  \right\vert \leq\frac{1}{8}M_{1}%
^{2}M_{2}^{3}\left\vert \lambda\right\vert ^{3}\text{ }.
\]

\end{lemma}

\begin{proof}
Since $f_{\overline{z}}^{-1}=\left(  I-E\right)  ^{-1}$ where $E=$ $\frac
{1}{2}\lambda J\mathcal{H}_{zz},$ for $\left\vert \lambda\right\vert
\leq\lambda_{\delta}\ $we have$\ \left\Vert E\right\Vert \leq\frac{1}{2}<1.$
By the Matrix Perturbation Lemma we have%
\begin{align*}
f_{\overline{z}}^{-1}  &  =I+E+E^{2}+E^{3}+E^{4}+\cdots\\
&  =I+E+E^{2}+E^{3}(I+E+\cdots)\\
&  =I+E+E^{2}+E^{3}f_{\overline{z}}^{-1}.
\end{align*}
Therefore,%
\begin{align}
\frac{\partial g(\lambda,z_{k})}{\partial\lambda}  &  =\mathcal{H}_{z}^{\top
}\overline{z}_{\lambda}\nonumber\\
&  =\frac{1}{2}\mathcal{H}_{z}^{\top}\left(  f_{\overline{z}}^{-1}J\right)
\mathcal{H}_{z}\nonumber\\
&  =\frac{1}{2}\left(  \mathcal{H}_{z}^{\top}J\mathcal{H}_{z}+\mathcal{H}%
_{z}^{\top}\left(  EJ\right)  \mathcal{H}_{z}\right. \nonumber\\
&  +\left.  \mathcal{H}_{z}^{\top}\left(  E^{2}J\right)  \mathcal{H}%
_{z}+\mathcal{H}_{z}^{\top}\left(  E^{3}f_{\overline{z}}^{-1}J\right)
\mathcal{H}_{z}\right)  . \label{dgdlam_expression}%
\end{align}
Since both $J$ and $E^{2}J=\frac{1}{4}\lambda^{2}\left(  J\mathcal{H}%
_{zz}J\mathcal{H}_{zz}J\right)  $ are skew-symmetric, the first and third term
in (\ref{dgdlam_expression}) equal zero. The second term is given by
\begin{align}
\mathcal{H}_{z}^{\top}\left(  EJ\right)  \mathcal{H}_{z}  &  =-\frac{1}%
{2}\lambda\left(  J\mathcal{H}_{z}\right)  ^{\top}\mathcal{H}_{zz}\left(
J\mathcal{H}_{z}\right) \nonumber\\
&  =-\frac{1}{2}\lambda\psi(\overline{z}(\lambda,z_{k}))\nonumber\\
&  =-\frac{1}{2}\lambda h(\lambda,z_{k}). \label{dgdlam_second_term}%
\end{align}
Thus, equations (\ref{dgdlam_expression}) and (\ref{dgdlam_second_term}) imply%
\begin{align*}
\left\vert \frac{\partial g(\lambda,z_{k})}{\partial\lambda}-\left(  -\frac
{1}{4}\lambda h(\lambda,z_{k})\right)  \right\vert  &  =\left\vert \frac{1}%
{2}\mathcal{H}_{z}^{\top}\left(  E^{3}f_{\overline{z}}^{-1}J\right)
\mathcal{H}_{z}\right\vert \\
&  \leq\frac{1}{8}M_{1}^{2}M_{2}^{3}\left\vert \lambda\right\vert ^{3}.
\end{align*}

\end{proof}

\begin{lemma}
\label{h_lambda}Assume $\psi\in C^{2}(U)$, $\left\Vert \psi_{z}(z)\right\Vert
\leq N_{1}$ and $\left\Vert \psi_{zz}(z)\right\Vert \leq N_{2}$ for $z\in U.$
Then, for $\lambda_{1},\lambda_{2}\in\lbrack-\mathcal{\lambda}_{\delta
},\mathcal{\lambda}_{\delta}]$ and $z_{k}\in U_{\delta},$
\[
\left\vert \frac{\partial h(\lambda_{2},z_{k})}{\partial\lambda}%
-\frac{\partial h(\lambda_{1},z_{k})}{\partial\lambda}\right\vert \leq
\gamma_{h}\left\vert \lambda_{2}-\lambda_{1}\right\vert \
\]
where $\gamma_{h}=N_{1}\gamma_{z}+M_{1}^{2}N_{2}.$
\end{lemma}

\noindent The proof is given in the appendix.

\begin{lemma}
\label{g_taylor}Assume $\psi\in C^{2}(U)$, $\left\Vert \psi_{z}(z)\right\Vert
\leq N_{1}$ and $\left\Vert \psi_{zz}(z)\right\Vert \leq N_{2}$ for $z\in U.$
Let $\mathcal{H}_{k}=\mathcal{H}(z_{k}),$ $\psi_{k}=\psi(z_{k})$ and $\psi
_{k}^{\prime}=\left.  [\psi,\mathcal{H}]\right\vert _{z=z_{k}}.$ Then, for
$\left\vert \lambda\right\vert \leq\lambda_{\delta}$ and $z_{k}\in U_{\delta
},$%
\begin{equation}
\left\vert \frac{\partial g(\lambda,z_{k})}{\partial\lambda}-\left(  -\frac
{1}{4}\psi_{k}\lambda-\frac{1}{8}\psi_{k}^{\prime}\,\lambda^{2}\right)
\right\vert \leq4K\left\vert \lambda\right\vert ^{3} \label{dgdlam_bound}%
\end{equation}
and%
\begin{equation}
\left\vert g(\lambda,z_{k})-\left(  \mathcal{H}_{k}-\frac{1}{8}\psi_{k}%
\lambda^{2}-\frac{1}{24}\psi_{k}^{\prime}\lambda^{3}\right)  \right\vert \leq
K\left\vert \lambda\right\vert ^{4} \label{g_bound}%
\end{equation}
where $K=\frac{1}{32}(M_{1}^{2}M_{2}^{3}+2\gamma_{h})$.
\end{lemma}

\begin{proof}
The Mean Value Theorem implies there exists a $\widetilde{\lambda}$ between
$0$ and $\lambda$ such that $h(\lambda,z_{k})-h(0,z_{k})=\left(  \partial
h(\widetilde{\lambda},z_{k})/\partial\lambda\right)  \lambda.$ Therefore,
using Lemma \ref{h_lambda},%
\begin{align*}
\left\vert h(\lambda,z_{k})-h(0,z_{k})-\frac{\partial h}{\partial\lambda
}(0,z_{k})\,\lambda\right\vert  &  =\left\vert \frac{\partial h}%
{\partial\lambda}(\widetilde{\lambda},z_{k})\,\lambda-\frac{\partial
h}{\partial\lambda}(0,z_{k})\,\lambda\right\vert \\
&  \leq\gamma_{h}|\widetilde{\lambda|}\left\vert \lambda\right\vert \leq
\gamma_{h}\left\vert \lambda\right\vert ^{2}.
\end{align*}
Since $h(0,z_{k})=\psi_{k}$ and $\partial h(0,z_{k})/\partial\lambda=\frac
{1}{2}\psi_{k}^{\prime},\ $we have $\left\vert h(\lambda,z_{k})-\psi_{k}%
-\frac{1}{2}\psi_{k}^{\prime}\,\lambda\right\vert \leq\gamma_{h}\left\vert
\lambda\right\vert ^{2}.$ Using Lemma \ref{g_lambda}, we establish inequality
(\ref{dgdlam_bound}) as follows.
\begin{align}
\left\vert \frac{\partial g(\lambda,z_{k})}{\partial\lambda}-\left(  -\frac
{1}{4}\psi_{k}\lambda-\frac{1}{8}\psi_{k}^{\prime}\,\lambda^{2}\right)
\right\vert  &  \leq\left\vert \frac{\partial g(\lambda,z_{k})}{\partial
\lambda}-\left(  -\frac{1}{4}\lambda h(\lambda,z_{k})\right)  \right\vert
\nonumber\\
&  +\left\vert \left(  -\frac{1}{4}\lambda h(\lambda,z_{k})\right)  -\left(
-\frac{1}{4}\psi_{k}\lambda-\frac{1}{8}\psi_{k}^{\prime}\,\lambda^{2}\right)
\right\vert \nonumber\\
&  \leq4K\left\vert \lambda\right\vert ^{3}\ \label{4K}%
\end{align}
where $K=\frac{1}{32}(M_{1}^{2}M_{2}^{3}+2\gamma_{h})$. We establish
(\ref{g_bound}) as follows. First, using (\ref{4K}) we have%
\begin{align*}
\left\vert
{\displaystyle\int_{0}^{\lambda}}
\left(  \frac{\partial g(\lambda,z_{k})}{\partial\lambda}+\frac{1}{4}\psi
_{k}\lambda+\frac{1}{8}\psi_{k}^{\prime}\,\lambda^{2}\right)  \,d\lambda
\right\vert  &  \leq%
{\displaystyle\int_{0}^{\lambda}}
\left\vert \frac{\partial g(\lambda,z_{k})}{\partial\lambda}+\frac{1}{4}%
\psi_{k}\lambda+\frac{1}{8}\psi_{k}^{\prime}\,\lambda^{2}\right\vert
\,d\lambda\\
&  \leq4K%
{\displaystyle\int_{0}^{\lambda}}
\left\vert \lambda\right\vert ^{3}\,d\lambda\\
&  =K\left\vert \lambda\right\vert ^{4}.
\end{align*}
But
\[%
{\displaystyle\int_{0}^{\lambda}}
\left(  \dfrac{\partial g(\lambda,z_{k})}{\partial\lambda}+\dfrac{1}{4}%
\psi_{k}\lambda+\dfrac{1}{8}\psi_{k}^{\prime}\,\lambda^{2}\right)
\,d\lambda=g(\lambda,z_{k})-g(0,z_{k})+\dfrac{1}{8}\psi_{k}\lambda^{2}%
+\dfrac{1}{24}\psi_{k}^{\prime}\,\lambda^{3}.
\]
So we have%
\[
\left\vert g(\lambda,z_{k})-\left(  \mathcal{H}_{k}-\frac{1}{8}\psi_{k}%
\lambda^{2}-\frac{1}{24}\psi_{k}^{\prime}\lambda^{3}\right)  \right\vert \leq
K\left\vert \lambda\right\vert ^{4}.
\]

\end{proof}

\section{Existence and Uniqueness of Lagrange Multipliers\label{multipliers}}

In this section, we address the question of the existence and uniqueness of
Lagrange multipliers $\lambda$ which satisfy the decoupled, Hamiltonian
constraint equation $g(\lambda,z_{k})=0.$ We begin by proving a monotonicity
result for the function $g(\lambda,z_{k}).$ Then we prove three separate
existence and uniqueness theorems, Theorems \ref{EU_1}--\ref{EU_3}, each of
which accounts for one of the three regions of extended-phase space described
below. (The value of the constant $K$ is determined by the Hamiltonian
function $\mathcal{H}(z).$ See Lemma \ref{g_taylor}.)%
\[%
\begin{tabular}
[c]{rl}%
region I & $\left\{  z_{k}:\psi(z_{k})\neq0,\ (\psi^{\prime}(z_{k}))^{2}%
\leq24K\left\vert \psi(z_{k})\right\vert \right\}  $\\
region II & $\left\{  z_{k}:\psi(z_{k})\neq0,\ (\psi^{\prime}(z_{k}%
))^{2}>24K\left\vert \psi(z_{k})\right\vert \right\}  $\\
region III & $\left\{  z_{k}:\psi(z_{k})=0,\ \psi^{\prime}(z_{k}%
)\neq0\right\}  .$%
\end{tabular}
\ \
\]

\begin{figure}[ptb]
\begin{center}
\subfigure[region I]{\label{bounds_a}\includegraphics
[height=1.32in]{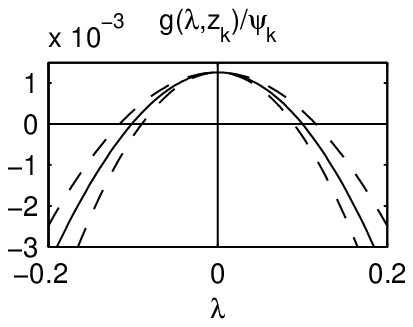}} \subfigure[region
II]{\label{bounds_b}\includegraphics
[height=1.32in]{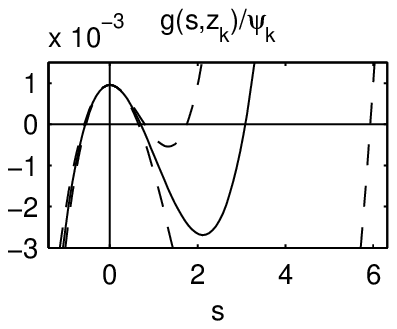}} \subfigure[region III
]{\label{bounds_c}\includegraphics [height=1.32in]{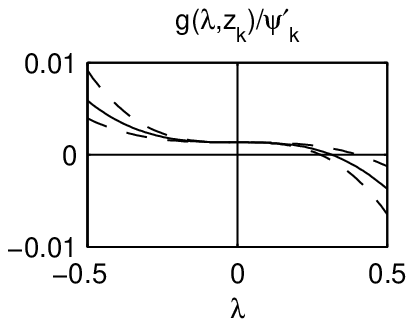}}
\end{center}
\caption{Bounds on $g(\lambda,z_{k})/\psi(z_{k})$ for the nonlinear pendulum.}%
\label{bounds}%
\end{figure}

The proofs of each of the three existence and uniqueness theorems in this
section uses the same basic approach. First, we derive bounds for the function
$g(\lambda,z_{k})/\psi(z_{k}).$ (See Figure \ref{bounds}.) Then, we use
monotonicity and the Intermediate Value Theorem to establish the existence and
(local) uniqueness of Lagrange multipliers $\lambda$ satisfying the equation
$g(\lambda,z_{k})=0$.

\begin{lemma}
[Monotonicity]\label{monotonicity}\ \ \ Assume $z_{k}\in U_{\delta}.$ Then we
claim the following:

\begin{enumerate}
\item[(i)] If $\psi_{k}\neq0,$ $(\psi_{k}^{\prime})^{2}\leq24K\left\vert
\psi_{k}\right\vert $ and $\left\vert \lambda\right\vert \leq\Lambda_{k}$
where $0<\Lambda_{k}<\min(\sqrt{\left\vert \psi_{k}\right\vert /96K}%
,\lambda_{\delta}),$ then $g(\lambda,z_{k})$ is monotonic
increasing/decreasing in the intervals $(-\Lambda_{k},0)$ and $(0,\Lambda
_{k}).$

\item[(ii)] Assume $\psi_{k}\neq0,$ $(\psi_{k}^{\prime})^{2}>24K\left\vert
\psi_{k}\right\vert $ and $\left\vert \lambda\right\vert \leq\Lambda_{k}$
where $0<\Lambda_{k}<\min(\left\vert \psi_{k}^{\prime}\right\vert
/48K,\lambda_{\delta}).$ Let $g(s,z_{k})$ be a reparametrization of
$g(\lambda,z_{k})$ where $s=-(\psi_{k}^{\prime}/\psi_{k})\lambda.$ Define
$S_{k}=\left\vert \psi_{k}^{\prime}/\psi_{k}\right\vert \Lambda_{k}$. Then
$g(s\,,z_{k})$ is monotonic increasing/decreasing in the following intervals:
a) $(-S_{k},0)$, b) $(0,S_{k})$ if $S_{k}<\frac{6}{5},$ c) $(0,\frac{6}{5})$
if $S_{k}\geq\frac{6}{5}$ and d) $(6,S_{k})$ if $S_{k}>6.$

\item[(iii)] If $\psi_{k}=0,$ $\psi_{k}^{\prime}\neq0,$ and $\left\vert
\lambda\right\vert \leq\Lambda_{k}$ where $0<\Lambda_{k}<\min(\left\vert
\psi_{k}^{\prime}\right\vert /48K,\lambda_{\delta}),$ then $g(\lambda,z_{k})$
is monotonic increasing/decreasing in the intervals $(-\Lambda_{k},0)$ and
$(0,\Lambda_{k}).$
\end{enumerate}
\end{lemma}

\noindent The proof of Lemma \ref{monotonicity} is given in the appendix.

\begin{figure}[ptb]
\begin{center}
\subfigure[$\mathcal{H}_{k}/\psi_{k}<0$]{\label{constraint_a1}\includegraphics
[height=1.32in]{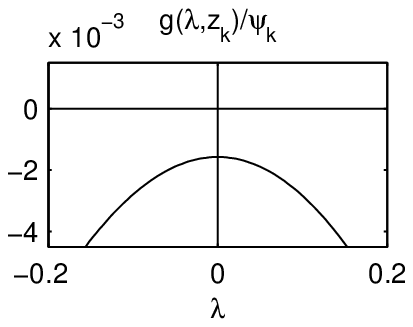}}
\subfigure[$\mathcal{H}_{k}/\psi_{k}=0$]{\label{constraint_a2}\includegraphics
[height=1.32in]{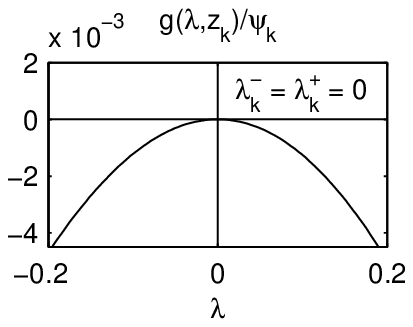}} \subfigure[$\mathcal{H}_{k}/\psi_{k}>0$
]{\label{constraint_a3}\includegraphics
[height=1.32in]{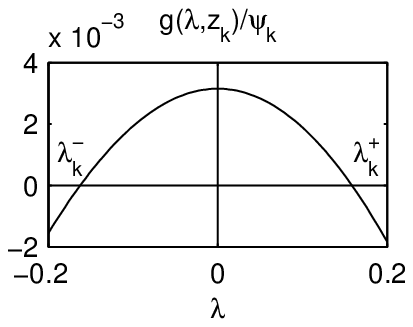}}
\end{center}
\caption{Plots of $g(\lambda,z_{k})/\psi(z_{k})$ in region I of the nonlinear
pendulum.}%
\label{constraint_a}%
\end{figure}

Theorem \ref{EU_1} below deals with region I of extended-phase space where the
quadratic term dominates the cubic term in the cubic approximation of
$g(\lambda,z_{k}).$ See Figure \ref{constraint_a} for plots of $g(\lambda
,z_{k})/\psi_{k}$ in region I of the nonlinear pendulum. Since $g(0,z_{k}%
)=\mathcal{H(}\overline{z}(0,z_{k}))=\mathcal{H}(z_{k})\mathcal{=H}_{k},$ we
see from Figure \ref{constraint_a} that the sign of $\mathcal{H}_{k}/\psi_{k}$
determines the number of solutions to the equation $g(\lambda,z_{k})=0.$

\begin{theorem}
\label{EU_1}Assume $z_{k}\in U_{\delta},$ $\psi_{k}\neq0$, $(\psi_{k}^{\prime
})^{2}\leq24K\left\vert \psi_{k}\right\vert \ $and $\left\vert \lambda
\right\vert <\Lambda_{k}$ where $0<\Lambda_{k}<\min(\sqrt{\left\vert \psi
_{k}\right\vert /96K},\lambda_{\delta}).$ Then the following statements about
the equation $g(\lambda,z_{k})=0$ are true$.$

\begin{enumerate}
\item[(i)] If $\mathcal{H}_{k}/\psi_{k}<0,$ no solution exists.

\item[(ii)] If $\mathcal{H}_{k}/\psi_{k}=0,$ the only solution is $\lambda=0$.

\item[(iii)] If $0<\mathcal{H}_{k}/\psi_{k}<\frac{3}{32}\Lambda_{k}^{2},$ two
solutions of opposite sign exist, $\lambda_{k}^{-}\in(-\Lambda_{k},0)$ and
$\lambda_{k}^{+}\in(0,\Lambda_{k})$. The solutions are unique within their
respective intervals.

\item[(iv)] If $\mathcal{H}_{k}/\psi_{k}>\frac{5}{32}\Lambda_{k}^{2},$ no
solution exists.
\end{enumerate}
\end{theorem}

\begin{proof}
Since $\left\vert \lambda\right\vert \leq\Lambda_{k}<\sqrt{\left\vert \psi
_{k}\right\vert /96K},$ we have from inequality (\ref{g_bound}) of Lemma
\ref{g_taylor} that
\[
\left\vert \frac{g(\lambda,z_{k})}{\psi_{k}}-\left(  \dfrac{\mathcal{H}_{k}%
}{\psi_{k}}-\frac{1}{8}\lambda^{2}-\frac{1}{24}\dfrac{\psi_{k}^{\prime}}%
{\psi_{k}}\lambda^{3}\right)  \right\vert \leq\frac{K}{\left\vert \psi
_{k}\right\vert }\left\vert \lambda\right\vert ^{4}\leq\frac{1}{96}\lambda
^{2}.
\]
It follows that%
\begin{equation}
\dfrac{\mathcal{H}_{k}}{\psi_{k}}-\frac{13}{96}\lambda^{2}-\frac{1}%
{24}\left\vert \dfrac{\psi_{k}^{\prime}}{\psi_{k}}\lambda\right\vert
\lambda^{2}\leq\frac{g(\lambda,z_{k})}{\psi_{k}}\leq\dfrac{\mathcal{H}_{k}%
}{\psi_{k}}-\frac{11}{96}\lambda^{2}+\frac{1}{24}\left\vert \dfrac{\psi
_{k}^{\prime}}{\psi_{k}}\lambda\right\vert \lambda^{2}. \label{g_bound_5}%
\end{equation}
Since by assumption $(\psi_{k}^{\prime})^{2}\leq24K\left\vert \psi
_{k}\right\vert ,$ we have
\begin{equation}
\left\vert \frac{\psi_{k}^{\prime}}{\psi_{k}}\lambda\right\vert =\left\vert
\frac{\psi_{k}^{\prime}}{\psi_{k}}\right\vert \left\vert \lambda\right\vert
<\frac{\sqrt{24K\left\vert \psi_{k}\right\vert }}{\left\vert \psi
_{k}\right\vert }\sqrt{\frac{\left\vert \psi_{k}\right\vert }{96K}}=\frac
{1}{2}. \label{psi_prime_bound}%
\end{equation}
Using (\ref{g_bound_5}) and (\ref{psi_prime_bound}) we have
\begin{equation}
\dfrac{\mathcal{H}_{k}}{\psi_{k}}-\frac{5}{32}\lambda^{2}\leq\frac
{g(\lambda,z_{k})}{\psi_{k}}\leq\dfrac{\mathcal{H}_{k}}{\psi_{k}}-\frac{3}%
{32}\lambda^{2}. \label{g_bound_1}%
\end{equation}
To establish (i), assume $\mathcal{H}_{k}/\psi_{k}<0.$ Then inequality
(\ref{g_bound_1}) implies $g(\lambda,z_{k})/\psi_{k}<0$ for all $\left\vert
\lambda\right\vert \leq\Lambda_{k}$ and no solution exists. If $\mathcal{H}%
_{k}/\psi_{k}=0,$ then $g(\lambda,z_{k})/\psi_{k}<0$ for nonzero $\lambda.$
Since $g(0,z_{k})=\mathcal{H}_{k}=0,$ the only solution is $\lambda=0,$
establishing (ii). If we assume $0<\mathcal{H}_{k}/\psi_{k}<\frac{3}%
{32}\Lambda_{k}^{2},$ then $g(\pm\Lambda_{k},z_{k})/\psi_{k}\leq
\mathcal{H}_{k}/\psi_{k}-\frac{3}{32}\Lambda_{k}^{2}<0.$ Since $g(0,z_{k}%
)/\psi_{k}=\mathcal{H}_{k}/\psi_{k}>0,$ the Intermediate Value Theorem implies
$g(\lambda,z_{k})=0$ has two solutions $\lambda_{k}^{-}\in(-\Lambda_{k},0)$
and $\lambda_{k}^{+}\in(0,\Lambda_{k}).$ Lemma \ref{monotonicity}(i) implies
$g(\lambda,z_{k})$ is monotonic in each interval establishing uniqueness and
claim (iii). Finally, if $\mathcal{H}_{k}/\psi_{k}>\frac{5}{32}\Lambda_{k}%
^{2},$ then inequality (\ref{g_bound_1}) implies that for all $\left\vert
\lambda\right\vert \leq\Lambda_{k},$
\[
0<\frac{\mathcal{H}_{k}}{\psi_{k}}-\frac{5}{32}\Lambda_{k}^{2}\leq
\frac{\mathcal{H}_{k}}{\psi_{k}}-\frac{5}{32}\lambda^{2}\leq\frac
{g(\lambda,z_{k})}{\psi_{k}}%
\]
establishing claim (iv).
\end{proof}

\begin{figure}[ptb]
\begin{center}
\subfigure[$\mathcal{H}_{k}/\psi_{k}<0$]{\label{constraint_b1}\includegraphics
[height=1.32in]{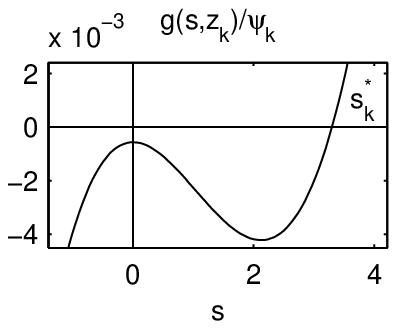}}
\subfigure[$\mathcal{H}_{k}/\psi_{k}=0$]{\label{constraint_b2}\includegraphics
[height=1.32in]{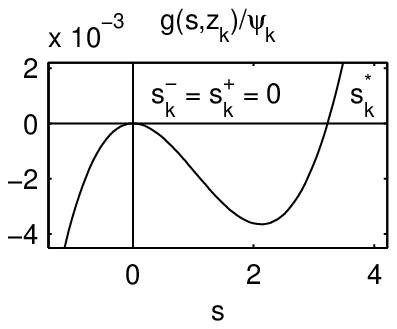}} \subfigure[$\mathcal{H}_{k}/\psi_{k}>0$
]{\label{constraint_b3}\includegraphics
[height=1.32in]{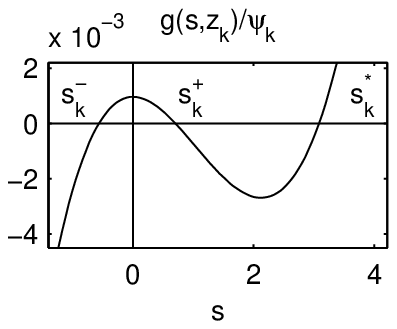}}
\end{center}
\caption{Plots of $g(\lambda,z_{k})/\psi(z_{k})$ in region II of the nonlinear
pendulum.}%
\label{constraint_b}%
\end{figure}

Theorem \ref{EU_2} below deals with region II\ of extended-phase space where
$\psi(z)$ is small but nonzero. Theorem \ref{EU_2} is the most complex of the
three existence and uniqueness theorems in this section because both quadratic
and cubic terms need to be taken into consideration. The reparametrization
$s=-\left(  \psi_{k}^{\prime}/\psi_{k}\right)  \lambda$ simplifies the
statement of the theorem and its proof. See Figure \ref{constraint_b} for
plots of $g(\lambda,z_{k})/\psi_{k}$ in region II of the nonlinear pendulum.
We can see from Figure \ref{constraint_b} that the sign of $\mathcal{H}%
_{k}/\psi_{k}$ determines the number of solutions to the equation
$g(s,z_{k})=0.$ (Recall that $g(0,z_{k})=\mathcal{H}_{k}.$) Note the
appearance of a \textquotedblleft ghost\ solution\textquotedblright,
$s_{k}^{\ast}$. From Figure \ref{constraint_b3}, we see that, unlike the two
solutions $s_{k}^{-}$ and $s_{k}^{+},$ the ghost solution $s_{k}^{\ast}$ does
not approach zero as $\mathcal{H}_{k}/\psi_{k}\rightarrow0^{+}.$

\begin{theorem}
\label{EU_2}Assume $z_{k}\in U_{\delta},$ $\psi_{k}\neq0$, $(\psi_{k}^{\prime
})^{2}>24K\left\vert \psi_{k}\right\vert \ $and $\left\vert \lambda\right\vert
<\Lambda_{k}$ where $0<\Lambda_{k}<\min(\left\vert \psi_{k}^{\prime
}\right\vert /48K,\lambda_{\delta}).$ Let $g(s,z_{k})$ be a reparametrization
of $g(\lambda,z_{k})$ where $s=-\left(  \psi_{k}^{\prime}/\psi_{k}\right)
\lambda$. Define $S_{k}=$ $\left\vert \psi_{k}^{\prime}/\psi_{k}\right\vert
\Lambda_{k}.$ Then the following statements about the equation $g(s,z_{k})=0$
are true$.$

\begin{enumerate}
\item[(i)] If $\mathcal{H}_{k}/\psi_{k}<0,$ no solution exists in the interval
$(-S_{k},2).$

\item[(ii)] If $\mathcal{H}_{k}/\psi_{k}=0,$ the only solution in the interval
$(-S_{k},2)$ is $s=0.$

\item[(iii)] If $\tfrac{1}{48}\Lambda_{k}^{2}(6-S_{k})<\mathcal{H}_{k}%
/\psi_{k}\leq0$ for $S_{k}>6,$ there exists a solution $s_{k}^{\ast}\in
\lbrack2,S_{k}).$

\item[(iv)] If $0<\mathcal{H}_{k}/\psi_{k}<\tfrac{1}{48}\Lambda_{k}%
^{2}(6+S_{k}),$ there exists a solution $s_{k}^{-}\in(-S_{k},0)\ $and the
solution is unique in this interval.

\item[(v)] If $0<\mathcal{H}_{k}/\psi_{k}<\tfrac{1}{16}\Lambda_{k}^{2}%
(2-S_{k})$ for $S_{k}<\frac{6}{5},$ there exists a solution $s_{k}^{+}%
\in(0,S_{k})$ and the solution is unique in this interval.

\item[(vi)] If $0\leq\mathcal{H}_{k}/\psi_{k}<\frac{9}{125}\left(  \psi
_{k}/\psi_{k}^{\prime}\right)  ^{2}$ and

\begin{enumerate}
\item if $S_{k}\geq\frac{6}{5},$ there exists a solution $s_{k}^{+}\in
\lbrack0,\frac{6}{5})$ and the solution is unique in this interval.

\item if $S_{k}\geq6,$ there exists a solution $s_{k}^{\ast}\in(\frac{6}%
{5},S_{k}).$
\end{enumerate}

\item[(vii)] If $\mathcal{H}_{k}/\psi_{k}>\frac{2}{3}\left(  \psi_{k}/\psi
_{k}^{\prime}\right)  ^{2},$ no solution exists in $(0,S_{k}).$
\end{enumerate}
\end{theorem}

\noindent The proof of Theorem \ref{EU_2} is given in the appendix.

\begin{figure}[ptb]
\begin{center}
\subfigure[$\mathcal{H}_{k}/\psi'_{k}<0$]{\label{constraint_c1}\includegraphics[height=1.32in]{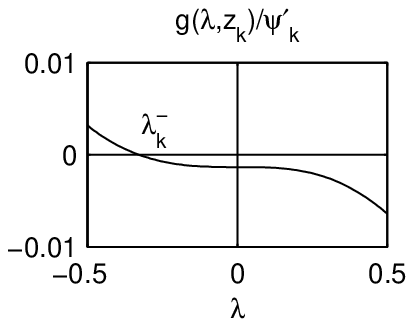}}
\subfigure[$\mathcal{H}_{k}/\psi'_{k}=0$]{\label{constraint_c2}\includegraphics[height=1.32in]{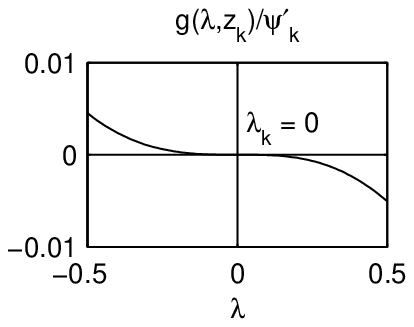}}
\subfigure[$\mathcal{H}_{k}/\psi'_{k}>0$
]{\label{constraint_c3}\includegraphics[height=1.32in]{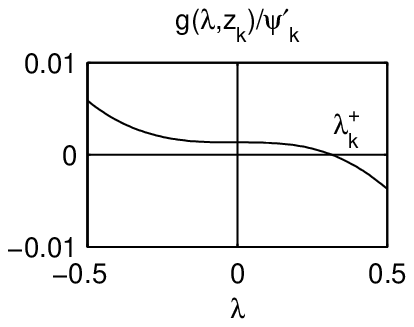}}
\end{center}
\caption{Plots of $g(\lambda,z_{k})/\psi\prime(z_{k})$ in region III of the
nonlinear pendulum.}%
\label{constraint_c}%
\end{figure}

Theorem \ref{EU_3} below deals with region III of extended-phase space where
the quadratic term of the cubic approximation of $g(\lambda,z_{k})$ is equal
to zero. See Figure \ref{constraint_c} for plots of $g(\lambda,z_{k})/\psi
_{k}^{\prime}$ in region III of the nonlinear pendulum. As we can see from
Figure \ref{constraint_c}, the sign of $\mathcal{H}_{k}/\psi_{k}^{\prime}$
determines whether the solution $\lambda_{k}^{-}$ or $\lambda_{k}^{+}$ exists.

\begin{theorem}
\label{EU_3}Assume $z_{k}\in U_{\delta},$ $\psi_{k}=0$, $\psi_{k}^{\prime}%
\neq0\ $and $\left\vert \lambda\right\vert \leq\Lambda_{k}$ where
$0<\Lambda_{k}<\min(\left\vert \psi_{k}^{\prime}\right\vert /48K,\lambda
_{\delta}).$ Then the following statements about the equation $g(\lambda
,z_{k})=0$ are true$.$

\begin{enumerate}
\item[(i)] If $-\tfrac{1}{48}\Lambda_{k}^{3}<\mathcal{H}_{k}/\psi_{k}^{\prime
}<0,$ there exists a solution $\lambda_{k}^{-}\in(-\Lambda_{k},0)$ and it is
unique in this interval. No solution exists in $[0,\Lambda_{k}).$

\item[(ii)] If $0<\mathcal{H}_{k}/\psi_{k}^{\prime}<\tfrac{1}{48}\Lambda
_{k}^{3},$ there exists a solution $\lambda_{k}^{+}\in(0,\Lambda_{k})$ and it
is unique in this interval. No solution exists in $(-\Lambda_{k},0].$

\item[(iii)] If $\mathcal{H}_{k}/\psi_{k}^{\prime}=0,$ the only solution is
$\lambda=0.$

\item[(iv)] If $\left\vert \mathcal{H}_{k}/\psi_{k}^{\prime}\right\vert
>\tfrac{1}{16}\Lambda_{k}^{3},$ no solution exists.
\end{enumerate}
\end{theorem}

\noindent The proof of Theorem \ref{EU_3} is given in the appendix.

\section{Existence and Uniqueness of DTH Trajectories \label{EU_section}}

The main result of this article is stated below in Theorem \ref{main_result}.
The proof of Theorem \ref{main_result} uses Theorems \ref{EU_1}--\ref{EU_3}
from the previous section. Before stating the theorem, we provide a condensed
description of the theorem's main conclusions.

\begin{figure}[ptb]
\begin{center}
\subfigure[passing through $z_0$.]{\label{DTH_trj_a}\includegraphics
[height=1.32in]{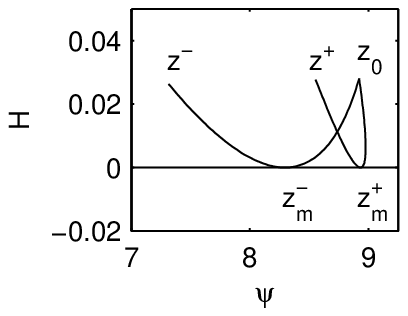}} \subfigure[bifurcating at
$z_0$.]{\label{DTH_trj_b}\includegraphics
[height=1.32in]{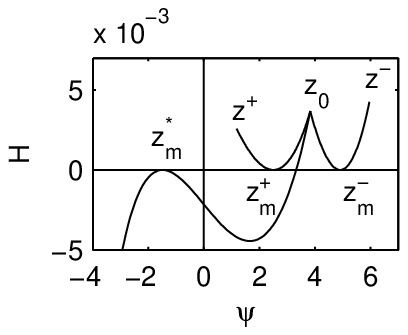}} \subfigure[terminating at
$z_0$.]{\label{DTH_trj_c}\includegraphics
[height=1.32in]{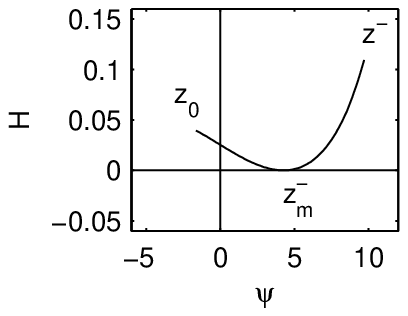}}
\end{center}
\caption{DTH trajectories of the nonlinear pendulum.}%
\label{DTH_trj}%
\end{figure}

Consider a point $z_{0}$ in extended-phase space. Roughly speaking, when
$\left\vert \mathcal{H}_{0}/\psi_{0}\right\vert $ is sufficiently small, there
are four generic possibilities for DTH\ trajectories. (1) If $\mathcal{H}%
_{0}/\psi_{0}\geq0$ and $\left\vert \psi_{0}\right\vert $ is large, then a
unique DTH trajectory exists which passes through the vertex point $z_{0.}$
(2) If $\mathcal{H}_{0}/\psi_{0}\geq0$ and $\psi$ changes sign near $z_{0}$,
then a DTH trajectory exists which bifurcates at the vertex point $z_{0.}$ (3)
If $\mathcal{H}_{0}/\psi_{0}<0$ and $\psi$ changes sign near $z_{0}$, then a
DTH trajectory exists which either begins or ends at the vertex point $z_{0}.$
(4) If $\mathcal{H}_{0}/\psi_{0}<0$ and $\left\vert \psi_{0}\right\vert $ is
large, then no DTH trajectory can exist having $z_{0}$ as a vertex point. See
Figure \ref{DTH_trj} for plots of DTH trajectories of the nonlinear pendulum
for the following initial conditions: (a) $z_{0}=(-0.15,-0.05,0,-3.48)$ (b)
$z_{0}=(1.25,-0.05,0,-4.33)$ (c) $z_{0}=(1.85,-0.05,0,-4.87)$.

Since DTH trajectories preserve the symplectic-energy-momentum properties of
Hamiltonian dynamics, Theorem \ref{main_result} provides conditions under
which a SEM integrator is well-posed. As a practical matter, we point out
that, for classical Hamiltonians, generic possibility (4), where no DTH
trajectory exists, can always be avoided by choosing an initial value for
$\wp_{0}\ $(the momentum conjugate to time) which is sufficiently small and of
the appropriate sign. Generic possibilities (2) and (3) are more challenging
to deal with and are discussed further in section \ref{ghost_trj}.

\begin{theorem}
[Existence \& Uniqueness of DTH Trajectories]\label{main_result}Consider an
extended-phase space Hamiltonian function $\mathcal{H}\in C^{2}(U)$ where
$U\subset\Re^{2n+2}$ is open. Define $\psi(z)=\left(  J\mathcal{H}_{z}\right)
^{\top}\mathcal{H}_{zz}\left(  J\mathcal{H}_{z}\right)  $ and $\psi^{\prime
}(z)=[\psi,\mathcal{H]}.$ Assume $\mathcal{H}_{z},$ $\mathcal{H}_{zz},$
$\psi_{z}$ and $\psi_{zz}$ are bounded on $U$ and $\mathcal{H}_{zz}$ is
Lipschitz continuous on $U.$ Assume also that $z_{k}\in U_{\delta}\ $where
$U_{\delta}\ =\left\{  z:\overline{B}(z,\delta)\subset U\right\}  $ and
$0<\delta<1$. If $\psi_{k}\neq0,$ then there exists a $\Lambda_{k}>0$ for
which statements (i)--(iii) are true.

\begin{enumerate}
\item[(i)] If $\mathcal{H}_{k}/\psi_{k}<0$, then $z_{k}$ can not be a vertex
point or end point of a DTH trajectory with Lagrange multiplier(s) $\left\vert
\lambda\right\vert \leq\Lambda_{k}$.

\item[(ii)] If $\mathcal{H}_{k}/\psi_{k}=0,$ then $z_{k}$ is a vertex point of
a fixed-point DTH trajectory with Lagrange multiplier $\lambda_{k}=0.$ No
other DTH trajectory with Lagrange multiplier(s) $\left\vert \lambda
_{k}\right\vert \leq\Lambda_{k}$ exists.

\item[(iii)] If $\mathcal{H}_{k}/\psi_{k}>0$ is sufficiently small, then
$z_{k}$ is a vertex point of a unique DTH trajectory passing through $z_{k}$
with Lagrange multipliers $\left\vert \lambda_{k}^{\pm}\right\vert \leq
\Lambda_{k}.$
\end{enumerate}

\noindent If $\left\vert \psi_{k}\right\vert \neq0$ is sufficiently small and
$\psi_{k}^{\prime}\neq0,$ then there exists a $\Lambda_{k}>0$ for which
statements (iv)--(vi) are true.

\begin{enumerate}
\item[(iv)] If $\mathcal{H}_{k}/\psi_{k}<0$ and $\left\vert \mathcal{H}%
_{k}/\psi_{k}\right\vert $ is sufficiently small, then $z_{k}$ is a vertex
point of a DTH trajectory which begins\ (ends) at $z_{k}$ and has Lagrange
multiplier $\left\vert \lambda_{k}^{\ast}\right\vert \leq\Lambda_{k}.$

\item[(v)] If $\mathcal{H}_{k}/\psi_{k}=0,$ then a DTH trajectory exists which
bifurcates at $z_{k}$ into a, fixed point, DTH trajectory with Lagrange
multiplier $\lambda_{k}=0$ and a ghost DTH trajectory with Lagrange multiplier
$\left\vert \lambda_{k}^{\ast}\right\vert \leq\Lambda_{k}.$

\item[(vi)] If $\mathcal{H}_{k}/\psi_{k}>0$ is sufficiently small, then a DTH
trajectory exists which bifurcates at $z_{k}$ into a DTH trajectory with
Lagrange multipliers $\left\vert \lambda_{k}^{\pm}\right\vert \leq\Lambda_{k}$
and a ghost DTH trajectory with Lagrange multiplier $\left\vert \lambda
_{k}^{\ast}\right\vert \leq\Lambda_{k}.$
\end{enumerate}

\noindent If $\psi_{k}=0$ and $\psi_{k}^{\prime}\neq0$, then there exists a
$\Lambda_{k}>0$ for which statements (vii) and (viii) are true.

\begin{enumerate}
\item[(vii)] If $\left\vert \mathcal{H}_{k}/\psi_{k}^{\prime}\right\vert
\neq0$ is sufficiently small, then $z_{k}$ is a vertex point of a unique DTH
trajectory which begins (ends) at $z_{k}$ with Lagrange multiplier $\left\vert
\lambda_{k}\right\vert \leq\Lambda_{k}.$

\item[(viii)] If $\mathcal{H}_{k}/\psi_{k}^{\prime}=0,$ then $z_{k}$ is a
vertex point of a fixed-point DTH trajectory with Lagrange multiplier
$\lambda_{k}=0.$ No other DTH trajectory with Lagrange multiplier $\left\vert
\lambda_{k}\right\vert \leq\Lambda_{k}$ exists.
\end{enumerate}
\end{theorem}

\begin{proof}
Consider the DTH equations
\begin{align}
\Delta z_{k}  &  =\lambda J\mathcal{H}_{z}(\overline{z})\label{DTH_eq_1}\\
\mathcal{H}(\overline{z})  &  =0 \label{DTH_eq_2}%
\end{align}
where $\Delta z_{k}=z-z_{k},$ $\overline{z}=\frac{1}{2}(z+z_{k})$ and
$\mathcal{H\in}C^{2}(U).$ We can rewrite equation (\ref{DTH_eq_1}) as follows:%
\begin{equation}
f(\lambda,z_{k},\overline{z})=\overline{z}-z_{k}-\frac{1}{2}\lambda
J\mathcal{H}_{z}(\overline{z})=0. \label{DTH_eq_1a}%
\end{equation}
By assumption, there exists $M_{1},\ M_{2}$ and $\gamma_{H}$ such that
$\left\Vert \mathcal{H}_{z}(z)\right\Vert \leq M_{1}$ and $\left\Vert
\mathcal{H}_{zz}(z)\right\Vert \leq M_{2}$ for $z\in U$ and $\left\Vert
\mathcal{H}_{zz}(z_{1})-\mathcal{H}_{zz}(z_{2})\right\Vert \leq\gamma
_{H}\left\Vert z_{1}-z_{2}\right\Vert $ for $z_{1},z_{2}\in U.$ Define%
\[
\mathcal{\lambda}_{\delta}=\min(\frac{1}{M_{2}},\frac{1}{\gamma_{H}}%
,\frac{1-(1-\delta)^{2}}{2M_{1}}).
\]
Theorem \ref{Decoupling} implies there exists a $0<\delta\,<1$ and a function
$\overline{z}(\lambda,z_{k})$ such that $f(\lambda,z_{k},\overline{z}%
(\lambda,z_{k}))=0$ for all $(\lambda,z_{k})\in\lbrack-\lambda_{\delta
},\lambda_{\delta}]\times U_{\delta}.$ Use the function $\overline{z}%
(\lambda,z_{k})$ to decouple equation (\ref{DTH_eq_2}) from equation
(\ref{DTH_eq_1}) to obtain the equation%
\begin{equation}
g(\lambda,z_{k})=\mathcal{H(}\overline{z}(\lambda,z_{k}))=0. \label{DTH_eq_1b}%
\end{equation}
For a given $z_{k}$ $\in U_{\delta},$ equation (\ref{DTH_eq_1b}) determines
the value of a Lagrange multiplier $\lambda_{k},$ provided that one exists. If
a Lagrange multiplier(s) exists, then $\lambda_{k}$ and $z_{k}$ determine
$z_{k-1}$ and/or $z_{k+1}$ as follows:%
\[
\left\{
\begin{tabular}
[c]{l}%
If $\lambda_{k}<0,$ define $\overline{z}_{k-1}=\overline{z}(\lambda_{k}%
,z_{k})\ $and $z_{k-1}=2\overline{z}_{k-1}-z_{k}.$\\
If $\lambda_{k}\geq0,$ define $\overline{z}_{k}=\overline{z}(\lambda_{k}%
,z_{k})\ $and $z_{k+1}=2\overline{z}_{k}-z_{k}.$%
\end{tabular}
\ \ \ \right.
\]
The extended-phase space, vertex points $z_{k-1},$ $z_{k}$ and $z_{k+1},$
determine a DTH trajectory which passes through the vertex point $z_{k}.$ If
only $z_{k-1}$ or $z_{k+1}$ exists, then the DTH trajectory either begins or
ends at $z_{k}.$

Now, we consider the existence and uniqueness of solutions to equation
(\ref{DTH_eq_1b}). By assumption, there exists $N_{1}$ and $N_{2}$ such that
$\left\Vert \psi_{z}(z)\right\Vert \leq N_{1}$ and $\left\Vert \psi
_{zz}(z)\right\Vert \leq N_{2}$ for $z\in U.$ Define
\[
K=\frac{1}{32}(M_{1}^{2}M_{2}^{3}+4M_{1}M_{2}N_{1}+2M_{1}^{2}N_{1}+2M_{1}%
^{2}N_{2}).
\]
Assume $\psi_{k}\neq0.$ If $(\psi_{k}^{\prime})^{2}\leq24K\left\vert \psi
_{k}\right\vert ,$ choose $0<\Lambda_{a}<\min(\sqrt{\left\vert \psi
_{k}\right\vert /96K},\lambda_{\delta}).$ Then Theorem \ref{EU_1} (i)--(iii)
imply statements (i)--(iii) are true. If, on the other hand, $(\psi
_{k}^{\prime})^{2}>24K\left\vert \psi_{k}\right\vert ,$ choose $0<\Lambda
_{b}<\min(\left\vert \psi_{k}^{\prime}\right\vert /48K,\lambda_{\delta}%
,\frac{6}{5}\left\vert \psi_{k}/\psi_{k}^{\prime}\right\vert ).$ For this
choice of $\Lambda_{b},$ we have $S_{k}=\left\vert \psi_{k}^{\prime}/\psi
_{k}\right\vert \Lambda_{b}<\left\vert \psi_{k}^{\prime}/\psi_{k}\right\vert
\frac{6}{5}\left\vert \psi_{k}/\psi_{k}^{\prime}\right\vert =\frac{6}{5}.$
Since $S_{k}<\frac{6}{5},$ Theorem \ref{EU_2} (i), (ii), (iv) and (v) imply
statements (i)--(iii). Therefore, for $\Lambda_{k}=$ $\min(\Lambda_{a}%
,\Lambda_{b}),$ statements (i)--(iii) are true.

Next, we prove (iv)--(vi). Assume $0<\left\vert \psi_{k}\right\vert
<\min((\psi_{k}^{\prime})^{2}/288K,\left\vert \psi_{k}^{\prime}\right\vert
\lambda_{\delta}/6).$ Then $6\left\vert \psi_{k}/\psi_{k}^{\prime}\right\vert
<$ $\min(\left\vert \psi_{k}^{\prime}\right\vert /48K,\lambda_{\delta})$.
Therefore, we can choose $\Lambda_{k}$ so that $6\left\vert \psi_{k}/\psi
_{k}^{\prime}\right\vert <$ $\Lambda_{k}<\min(\left\vert \psi_{k}^{\prime
}\right\vert /48K,\lambda_{\delta}).$ Then we have $S_{k}=\left\vert \psi
_{k}^{\prime}/\psi_{k}\right\vert \Lambda_{k}>\left\vert \psi_{k}^{\prime
}/\psi_{k}\right\vert 6\left\vert \psi_{k}/\psi_{k}^{\prime}\right\vert =6.$
If $\mathcal{H}_{k}/\psi_{k}<0,$ Theorem \ref{EU_2}(i) implies no solution
exists in $(-S_{k},2).$ Since $S_{k}>6,$ Theorem \ref{EU_2}(iii) implies that
for sufficiently small $\left\vert \mathcal{H}_{k}/\psi_{k}\right\vert ,$ a
solution $s_{k}^{\ast}\in\lbrack2,S_{k})$ exists. Since $\lambda_{k}^{\ast
}=-(\psi_{k}/\psi_{k}^{\prime})s_{k}^{\ast},$ and $s_{k}^{\ast}>0,$ all
solutions that may exist must have the same sign. Hence, the DTH trajectory
must begin (end) at $z_{k}$, proving statement (iv). If $\mathcal{H}_{k}%
/\psi_{k}=0$, Theorem \ref{EU_2}(ii), (vi)b prove statement (v). If
$\mathcal{H}_{k}/\psi_{k}>0$ is sufficiently small, Theorem \ref{EU_2}(iv),
(vi) prove statement (vi).

Finally, assume $\psi_{k}=0$ and $\psi_{k}^{\prime}\neq0.$ Choose
$0<\Lambda_{k}<\min(\left\vert \psi_{k}^{\prime}\right\vert /48K,\lambda
_{\delta}).$ Theorem \ref{EU_3}(i)--(iii), imply (vii)--(viii).
\end{proof}

\section{Ghost Trajectories \label{ghost_trj}}

Discrete approximations of differential equations can introduce spurious or
nonphysical solutions. Greenspan \cite{Greenspan-84} provided a detailed
analysis of a \textquotedblleft nonphysical\textquotedblright\ solution to his
equations for discrete mechanics. Greenspan showed that, unlike the correct
physical solution, the nonphysical solution approaches infinity as the time
step is brought to zero.

Multiple solutions also exist in DTH dynamics. When $\left\vert \psi
(z)\right\vert $ is large, the decoupled Hamiltonian constraint equation
$g(\lambda,z_{k})=0$ has only two solutions, $\lambda_{k}^{-}$ and
$\lambda_{k}^{+},$ both of which appear to represent the correct physical
behavior of the system. The Lagrange multiplier $\lambda_{k}^{-}$ corresponds
to the trajectory propagating backward in time from $z_{k}$ and $\lambda
_{k}^{+}$ corresponds to the trajectory propagating forward in time from
$z_{k}.$

Near points where $\psi(z)$ changes sign, a third solution to $g(\lambda
,z_{k})=0$ appears---the solution $\lambda_{k}^{\ast}.$ As stated earlier, the
solution $\lambda_{k}^{\ast}$ has a property that distinguishes it from the
solutions $\lambda_{k}^{-}$ and $\lambda_{k}^{+}.$ Assume a sequence of
initial conditions $z_{k}$ approaches the Hamiltonian conserving manifold
$\mathcal{H}(z)=0,$ but not the manifold $\psi(z)=0.$ Then the corresponding
sequences, $\lambda_{k}^{-}$ and $\lambda_{k}^{+},$ each converge to zero, but
the sequence $\lambda_{k}^{\ast}$ does \emph{not} converge to zero. We make
this property precise in Theorem \ref{ghost} below.

The solution $\lambda_{k}^{\ast}$ causes a DTH trajectory to bifurcate at
$z_{k},$ giving rise to what we call \textquotedblleft ghost\textquotedblright%
\ DTH trajectories. Ghost DTH trajectories are not time reversible. (See
Shibberu \cite{Shibberu-05} for the details.) We will refrain from calling
ghost trajectories \textquotedblleft nonphysical\textquotedblright\ because,
in DTH dynamics, it is unclear what the physically correct solution across
$\psi(z)=0$ manifolds should be. It appears that DTH dynamics needs to be
regularized in some fashion. In Shibberu \cite{Shibberu-05}, we propose a
regularization of DTH dynamics which preserves symplectic-energy-momentum
properties and time reversibility across $\psi(z)=0$ manifolds.

\begin{theorem}
\label{ghost}Consider a sequence $z_{k}\in U_{\delta},$ $k=0,1,2,\ldots$ where
$\left\vert \psi(z_{k})\right\vert >\psi_{\min}>0$ and $\mathcal{H}%
(z_{k})/\psi(z_{k})\rightarrow0^{+}.$ If $\lambda_{k}^{\pm}$ exist, then
$\lim_{k\rightarrow\infty}\lambda_{k}^{\pm}=0.$ If $\lim_{k\rightarrow\infty
}\lambda_{k}^{\ast}$ exists, then $\lim_{k\rightarrow\infty}\left\vert
\lambda_{k}^{\ast}\right\vert >\frac{6}{5}\left(  \psi_{\min}/M_{1}%
N_{1}\right)  >0.$
\end{theorem}

\begin{proof}
Assume $z_{k}$ is in region I of extended-phase space. Then, inequality
(\ref{g_bound_1}) of Theorem \ref{EU_1} implies%
\begin{equation}
0\leq\frac{3}{32}\left(  \lambda_{k}^{\pm}\right)  ^{2}\leq\frac
{\mathcal{H}(z_{k})}{\psi(z_{k})}. \label{ghost_a}%
\end{equation}
Now assume $z_{k}$ is in region II. Depending on the sign of $\psi_{k}%
^{\prime}/\psi_{k},$ either $s_{k}^{-}=-\left(  \psi_{k}^{\prime}/\psi
_{k,}\right)  \lambda_{k}^{-}$ or $s_{k}^{-}=-\left(  \psi_{k}^{\prime}%
/\psi_{k,}\right)  \lambda_{k}^{+}.$ Therefore, inequality (\ref{g_bound_2a})
of Theorem \ref{EU_2} implies either%
\begin{equation}
0\leq\frac{1}{48}\left(  \lambda_{k}^{+}\right)  ^{2}\left(  6+\left\vert
s_{k}^{-}\right\vert \right)  \leq\frac{\mathcal{H}(z_{k})}{\psi(z_{k})}
\label{ghost_b1}%
\end{equation}
or%
\begin{equation}
0\leq\frac{1}{48}\left(  \lambda_{k}^{-}\right)  ^{2}\left(  6+\left\vert
s_{k}^{-}\right\vert \right)  \leq\frac{\mathcal{H}(z_{k})}{\psi(z_{k})}
\label{ghost_b2}%
\end{equation}
Likewise, since $s_{k}^{+}\leq\frac{6}{5}<2,$ inequality (\ref{g_bound_2b})
implies, in correspondence with inequalities (\ref{ghost_b1})--(\ref{ghost_b2}%
), either%
\begin{equation}
0\leq\frac{1}{16}\left(  \lambda_{k}^{-}\right)  ^{2}\left(  2-s_{k}%
^{+}\right)  \leq\frac{\mathcal{H}(z_{k})}{\psi(z_{k})} \label{ghost_c1}%
\end{equation}
or%
\begin{equation}
0\leq\frac{1}{16}\left(  \lambda_{k}^{+}\right)  ^{2}\left(  2-s_{k}%
^{+}\right)  \leq\frac{\mathcal{H}(z_{k})}{\psi(z_{k})}. \label{ghost_c2}%
\end{equation}
Since $\mathcal{H}_{k}/\psi_{k}\rightarrow0^{+},$ inequalities (\ref{ghost_a}%
)--(\ref{ghost_c2}) imply $\lim_{k\rightarrow\infty}\lambda_{k}^{\pm}=0.$
Since $\left\vert s_{k}^{\ast}\right\vert =\left\vert -(\psi_{k}^{\prime}%
/\psi_{k})\lambda_{k}^{\ast}\right\vert >\frac{6}{5},$ we have $\left\vert
\lambda_{k}^{\ast}\right\vert >\frac{6}{5}\left\vert \psi_{k}/\psi_{k}%
^{\prime}\right\vert >\frac{6}{5}\left(  \psi_{\min}/M_{1}N_{1}\right)  .$ If
$\lim_{k\rightarrow\infty}\lambda_{k}^{\ast}$ exists, we must have
$\lim_{k\rightarrow\infty}\left\vert \lambda_{k}^{\ast}\right\vert >\frac
{6}{5}\left(  \psi_{\min}/M_{1}N_{1}\right)  >0.$
\end{proof}

\section{Conclusions}

The extended-phase space formulation of the principle of least action leads to
indeterminate equations of motion. Since SEM integration is based on a
discrete version of this principle, it is important to establish conditions
under which the equations of SEM integration are well-posed. Theorem
\ref{main_result} provides such conditions. Theorem \ref{main_result} also
shows that the DTH equations of SEM integration need to be regularized in some
fashion. One proposal for regularizing SEM integration is given in Shibberu
\cite{Shibberu-05}.

The existence and uniqueness results in this article are only locally valid. A
global result---for example, sufficient conditions for the existence of DTH
trajectories for arbitrarily long intervals of time---would be interesting.
One of the difficulties in establishing such a result appears to be
establishing a global bound on the Lagrange multipliers $\lambda_{k},$
$k=0,1,\ldots$.

A coordinate-invariant, formulation of DTH dynamics could provide additional
insight into the behavior of DTH trajectories crossing $\psi(z)=0$ manifolds.
Preliminary work on a coordinate-invariant formulation of DTH was given in
Shibberu \cite{Shibberu-92}. The mathematical tools developed and refined in
Talasila, Clemente-Gallardo, van der Schaft \cite{Talasila-04} and Desbrun,
Hirani, Leok and Marsden \cite{Desbrun-05} could prove useful in developing a
more rigorous, coordinate-invariant formulation of DTH dynamics and SEM integration.

\section*{Acknowledgements}

The author wishes to acknowledge a helpful discussion he had with Raymond
Chin, IUPUI, concerning cubic approximations. The author also thanks his
siblings, Hebret, Saba and Dagmawy for their encouragement and support.

\section*{Appendix}

\subsection*{Lemma \ref{z_lambda}}

\begin{proof}
Using implicit differentiation, we have $\left\Vert \overline{z}_{\lambda
}\right\Vert =\left\Vert f_{\overline{z}}^{-1}\frac{1}{2}J\mathcal{H}%
_{z}\right\Vert \leq M_{1}.$ Therefore $\left\Vert \overline{z}(\lambda
_{2},z_{k})-\overline{z}(\lambda_{1},z_{k})\right\Vert \leq M_{1}\left\vert
\lambda_{2}-\lambda_{1}\right\vert .$ Using the abbreviation $f_{\overline{z}%
}(\lambda)$ for $f_{\overline{z}}(\lambda,z_{k},\overline{z}(\lambda,z_{k}))$
we have%
\begin{align*}
&  \left\Vert f_{\overline{z}}(\lambda_{2})-f_{\overline{z}}(\lambda
_{1})\right\Vert =\frac{1}{2}\left\Vert \lambda_{2}J\mathcal{H}_{zz}%
(\overline{z}_{2})-\lambda_{1}J\mathcal{H}_{zz}(\overline{z}_{1})\right\Vert
\\
&  \leq\frac{1}{2}\left\vert \lambda_{2}\right\vert \left\Vert \mathcal{H}%
_{zz}(\overline{z}_{2})-\mathcal{H}_{zz}(\overline{z}_{1})\right\Vert
+\frac{1}{2}\left\Vert \mathcal{H}_{zz}(\overline{z}_{1})\right\Vert
\left\vert \lambda_{2}-\lambda_{1}\right\vert \\
&  \leq\frac{1}{2}\lambda_{\delta}\gamma_{H}\left\Vert \overline{z}%
_{2}-\overline{z}_{1}\right\Vert +\frac{1}{2}M_{2}\left\vert \lambda
_{2}-\lambda_{1}\right\vert \\
&  \leq\frac{1}{2}\left(  \lambda_{\delta}\gamma_{H}\,M_{1}+M_{2}\right)
\left\vert \lambda_{2}-\lambda_{1}\right\vert .
\end{align*}
Therefore,%
\begin{align*}
&  \left\Vert \overline{z}_{\lambda}(\lambda_{2},z_{k})-\overline{z}_{\lambda
}(\lambda_{1},z_{k})\right\Vert =\frac{1}{2}\left\Vert f_{\overline{z}}%
^{-1}(\lambda_{2})J\mathcal{H}_{z}\left(  \overline{z}_{2}\right)
-f_{\overline{z}}^{-1}(\lambda_{1})J\mathcal{H}_{z}\left(  \overline{z}%
_{1}\right)  \right\Vert \\
&  \leq\frac{1}{2}\left\Vert f_{\overline{z}}^{-1}(\lambda_{2})\right\Vert
\left\Vert \mathcal{H}_{z}\left(  \overline{z}_{2}\right)  -\mathcal{H}%
_{z}\left(  \overline{z}_{1}\right)  \right\Vert +\frac{1}{2}\left\Vert
\mathcal{H}_{z}\left(  \overline{z}_{1}\right)  \right\Vert \left\Vert
f_{\overline{z}}^{-1}(\lambda_{2})-f_{\overline{z}}^{-1}(\lambda
_{1})\right\Vert \\
&  \leq M_{2}\left\Vert \overline{z}_{2}-\overline{z}_{1}\right\Vert +\frac
{1}{2}M_{1}\left\Vert f_{\overline{z}}^{-1}(\lambda_{2})\right\Vert \left\Vert
f_{\overline{z}}(\lambda_{2})-f_{\overline{z}}(\lambda_{1})\right\Vert
\left\Vert f_{\overline{z}}^{-1}(\lambda_{1})\right\Vert \\
&  \leq M_{2}M_{1}\left\vert \lambda_{2}-\lambda_{1}\right\vert +M_{1}\left(
\lambda_{\delta}\gamma_{H}\,M_{1}+M_{2}\right)  \left\vert \lambda_{2}%
-\lambda_{1}\right\vert \\
&  \leq\left(  2M_{1}M_{2}+\,M_{1}^{2}\right)  \left\vert \lambda_{2}%
-\lambda_{1}\right\vert \\
&  =\gamma_{z}\left\vert \lambda_{2}-\lambda_{1}\right\vert .
\end{align*}

\end{proof}

\subsection*{Lemma \ref{h_lambda}}

\begin{proof}
Using Lemma \ref{z_lambda} and the fact that $h(\lambda,z_{k})=\psi
\mathcal{(}\overline{z}(\lambda,z_{k}))$ we have
\begin{align*}
\left\vert \dfrac{\partial h(\lambda_{2},z_{k})}{\partial\lambda}%
-\dfrac{\partial h(\lambda_{1},z_{k})}{\partial\lambda}\right\vert  &
=\left\vert \psi_{z}^{\top}(\overline{z}_{2})\overline{z}_{\lambda}%
(\lambda_{2},z_{k})-\psi_{z}^{\top}(\overline{z}_{1})\overline{z}_{\lambda
}(\lambda_{1},z_{k})\right\vert \\
&  \leq\left\Vert \psi_{z}^{\top}(\overline{z}_{2})\right\Vert \left\Vert
\overline{z}_{\lambda}(\lambda_{2},z_{k})-\overline{z}_{\lambda}(\lambda
_{1},z_{k})\right\Vert \\
&  +\left\Vert \overline{z}_{\lambda}(\lambda_{1},z_{k})\right\Vert \left\Vert
\psi_{z}^{\top}(\overline{z}_{2})-\psi_{z}^{\top}(\overline{z}_{1})\right\Vert
\\
&  \leq\left(  N_{1}\gamma_{z}+M_{1}^{2}N_{2}\right)  \left\vert \lambda
_{2}-\lambda_{1}\right\vert \\
&  =\gamma_{h}\left\vert \lambda_{2}-\lambda_{1}\right\vert .
\end{align*}

\end{proof}

\subsection*{Lemma \ref{monotonicity} (Monotonicity)}

\begin{proof}
Inequality (\ref{dgdlam_bound}) of Lemma \ref{g_taylor} implies%
\begin{equation}
\left\vert \frac{\partial g}{\partial\lambda}+\frac{1}{4}\lambda\left(
\psi_{k}+\frac{1}{2}\psi_{k}^{\prime}\lambda\right)  \right\vert
\leq4K\left\vert \lambda\right\vert ^{3}. \label{mono_a}%
\end{equation}
Under the assumptions of claim (i), we have that $\left\vert \frac{1}{2}%
\psi_{k}^{\prime}\lambda\right\vert =\frac{1}{2}\left\vert \psi_{k}^{\prime
}\right\vert \left\vert \lambda\right\vert \leq\sqrt{24K\left\vert \psi
_{k}\right\vert }\sqrt{\left\vert \psi_{k}\right\vert /96K}=\frac{1}%
{4}\left\vert \psi_{k}\right\vert .$ Therefore, $\frac{1}{4}\left\vert
\psi_{k}\right\vert >\left\vert \frac{1}{2}\psi_{k}^{\prime}\lambda\right\vert
,$ and we have%
\begin{equation}
\frac{3}{4}\left\vert \psi_{k}\right\vert =\left\vert \psi_{k}\right\vert
-\frac{1}{4}\left\vert \psi_{k}\right\vert \leq\left\vert \psi_{k}+\frac{1}%
{2}\psi_{k}^{\prime}\lambda\right\vert . \label{mono_b}%
\end{equation}
Now, assume $\partial g/\partial\lambda=0$ for $\lambda\neq0.$ Then
(\ref{mono_a}) implies $\frac{1}{4}\left\vert \lambda\right\vert \left\vert
\psi_{k}+\frac{1}{2}\psi_{k}^{\prime}\lambda\right\vert \leq4K\left\vert
\lambda\right\vert ^{3}<4K\left(  \left\vert \psi_{k}\right\vert /96K\right)
\left\vert \lambda\right\vert =\frac{1}{24}\left\vert \psi_{k}\right\vert
\left\vert \lambda\right\vert .$ Therefore,%
\begin{equation}
\left\vert \psi_{k}+\frac{1}{2}\psi_{k}^{\prime}\lambda\right\vert \leq
\frac{1}{6}\left\vert \psi_{k}\right\vert . \label{mono_c}%
\end{equation}
Inequalities (\ref{mono_b}) and (\ref{mono_c}) imply $\frac{3}{4}\leq\frac
{1}{6},$ a contradiction. Therefore $\partial g/\partial\lambda\neq0$ for
$\lambda\neq0$ and hence $g(\lambda,z_{k})$ is monotonic increasing/decreasing
on the intervals $(-\Lambda_{k},0)$ and $(0,\Lambda_{k})$.

Under the assumptions of claim (ii), inequality (\ref{mono_a}) becomes%
\begin{equation}
\left\vert \frac{\partial g(s,z_{k})/\partial\lambda}{\psi_{k}}-\frac{1}%
{8}\frac{\psi_{k}}{\psi_{k}^{\prime}}s\left(  2-s\right)  \right\vert
\leq\frac{1}{12}\left\vert \frac{\psi_{k}}{\psi_{k}^{\prime}}\right\vert
s^{2}. \label{mono_d}%
\end{equation}
Assume $\partial g(s,z_{k})/\partial\lambda=0$ for $s\neq0.$ Then
(\ref{mono_d}) becomes%
\begin{equation}
\left\vert 2-s\right\vert \leq\frac{2}{3}\left\vert s\right\vert .
\label{mono_e}%
\end{equation}
If $s<0,$ (\ref{mono_e}) implies $2-s\leq-\frac{2}{3}s$ or $s\geq6,$ a
contradiction. If $0<s\leq2,$ then (\ref{mono_e}) implies $2-s\leq\frac{2}%
{3}s$ or $s\geq\frac{6}{5}.$ If $s>2,$ then we have $s-2\leq\frac{2}{3}s$ or
$s\leq6.$ Thus, $\partial g(s,z_{k})/\partial\lambda$ can equal zero only if
$s=0$ or $\frac{6}{5}\leq s\leq6.$ Therefore, $g(s,z_{k})$ is monotonic
increasing/decreasing in the interval $(-S_{k},0)$, in the interval
$(0,S_{k})$ if $S_{k}<\frac{6}{5},$ in the interval $(0,\frac{6}{5})$ if
$S_{k}\geq\frac{6}{5}$ and in the interval $(6,S_{k})$ if $S_{k}>6.$

Finally, under the assumptions of claim (iii), if $\partial g/\partial
\lambda=0$ for $\lambda\neq0,$ inequality (\ref{mono_a}) becomes $\frac{1}%
{8}\left\vert \psi_{k}^{\prime}\right\vert \lambda^{2}\leq4K\left\vert
\lambda\right\vert ^{3}\leq4K(\left\vert \psi_{k}^{\prime}\right\vert
/48K)\lambda^{2}=\frac{1}{12}\left\vert \psi_{k}^{\prime}\right\vert
\lambda^{2}$ which implies $\frac{1}{8}\leq\frac{1}{12},$ a contradiction.
Therefore $\partial g/\partial\lambda\ $is nonzero for $\lambda\neq0$ and
$g(\lambda,z_{k})$ is monotonic increasing/decreasing in the intervals
$(-\Lambda_{k},0)$ and $(0,\Lambda_{k}).$
\end{proof}

\subsection*{Theorem \ref{EU_2}}

\begin{proof}
We have from equation (\ref{g_bound}) of Lemma \ref{g_taylor} that
\[
\left\vert \frac{g(\lambda,z_{k})}{\psi_{k}}-\left(  \dfrac{\mathcal{H}_{k}%
}{\psi_{k}}-\frac{1}{8}\lambda^{2}-\frac{1}{24}\dfrac{\psi_{k}^{\prime}}%
{\psi_{k}}\lambda^{3}\right)  \right\vert \leq\frac{K}{\left\vert \psi
_{k}\right\vert }\left\vert \lambda\right\vert ^{4}\leq\frac{1}{48}\left\vert
\dfrac{\psi_{k}^{\prime}}{\psi_{k}}\right\vert \left\vert \lambda\right\vert
^{3}.
\]
Using the reparametrization $\lambda=-\left(  \psi_{k}/\psi_{k}^{\prime
}\right)  s$ we have%
\[
\left\vert \frac{g(s,z_{k})}{\psi_{k}}-\left(  \dfrac{\mathcal{H}_{k}}%
{\psi_{k}}-\frac{1}{8}\left(  \dfrac{\psi_{k}}{\psi_{k}^{\prime}}\right)
^{2}s^{2}+\frac{1}{24}\left(  \dfrac{\psi_{k}}{\psi_{k}^{\prime}}\right)
^{2}s^{3}\right)  \right\vert \leq\frac{1}{48}\left(  \dfrac{\psi_{k}}%
{\psi_{k}^{\prime}}\right)  ^{2}\left\vert s\right\vert ^{3}.
\]
For $-S_{k}\leq s\leq0$ we have%
\begin{equation}
\dfrac{\mathcal{H}_{k}}{\psi_{k}}-\frac{1}{16}\lambda^{2}(2-s)\leq
\frac{g(s,z_{k})}{\psi_{k}}\leq\dfrac{\mathcal{H}_{k}}{\psi_{k}}-\frac{1}%
{48}\lambda^{2}(6-s). \label{g_bound_2a}%
\end{equation}
For $0\leq s\leq S_{k}$ we have%
\begin{equation}
\dfrac{\mathcal{H}_{k}}{\psi_{k}}-\frac{1}{48}\lambda^{2}(6-s)\leq
\frac{g(s,z_{k})}{\psi_{k}}\leq\dfrac{\mathcal{H}_{k}}{\psi_{k}}-\frac{1}%
{16}\lambda^{2}(2-s). \label{g_bound_2b}%
\end{equation}
Now, by inequality (\ref{g_bound_2a}), if $-S_{k}\leq s<0$ and $\mathcal{H}%
_{k}/\psi_{k}\leq0,$ then%
\begin{equation}
\frac{g(s,z_{k})}{\psi_{k}}\leq-\left\vert \dfrac{\mathcal{H}_{k}}{\psi_{k}%
}\right\vert -\frac{1}{48}\lambda^{2}(6+\left\vert s\right\vert )<0.
\label{g_bound_2c}%
\end{equation}
Similarly, by inequality (\ref{g_bound_2b}), if $0<s<2$ and $\mathcal{H}%
_{k}/\psi_{k}\leq0,$ then%
\begin{equation}
\frac{g(s,z_{k})}{\psi_{k}}\leq-\left\vert \dfrac{\mathcal{H}_{k}}{\psi_{k}%
}\right\vert -\frac{1}{16}\lambda^{2}(2-\left\vert s\right\vert )<0.
\label{g_bound_2d}%
\end{equation}
Since $g(0,z_{k})/\psi_{k}=\mathcal{H}_{k}/\psi_{k},$ inequality
(\ref{g_bound_2c}) and (\ref{g_bound_2d}) imply the following: If
$\mathcal{H}_{k}/\psi_{k}$ is strictly less than zero, then no solution exists
in the interval $(-S_{k},2)$ establishing claim (i). If $\mathcal{H}_{k}%
/\psi_{k}$ equals zero, then $s=0$ is the only solution in $(-S_{k},2)$
establishing claim (ii).

Next, we use the Intermediate Value Theorem to establish claim (iii).
Inequality (\ref{g_bound_2d}) implies that for $0<s_{o}<2,$%
\begin{equation}
\frac{g(s_{o},z_{k})}{\psi_{k}}\leq-\left\vert \dfrac{\mathcal{H}_{k}}%
{\psi_{k}}\right\vert -\dfrac{1}{16}\lambda_{o}^{2}(2-s_{o})<0.
\label{g_bound_2e}%
\end{equation}
Assume $S_{k}>6$ and $\frac{1}{48}\Lambda_{k}(6-S_{k})<\mathcal{H}_{k}%
/\psi_{k}\leq0.$ Using inequality (\ref{g_bound_2b}) we have%
\begin{equation}
0<\dfrac{\mathcal{H}_{k}}{\psi_{k}}-\dfrac{1}{48}\Lambda_{k}^{2}(6-S_{k}%
)\leq\frac{g(S_{k},z_{k})}{\psi_{k}}. \label{g_bound_2f}%
\end{equation}
Inequality (\ref{g_bound_2e}) and (\ref{g_bound_2f}) and the Intermediate
Value Theorem imply there must exist a solution $s_{k}^{\ast}\in\lbrack
2,S_{k})$ establishing claim (iii).

Proceeding in a similar fashion, if $0<\mathcal{H}_{k}/\psi_{k}<\frac{1}%
{48}\Lambda_{k}^{2}(6+S_{k})$ and $-S_{k}<s<0,$ inequality (\ref{g_bound_2a})
implies%
\[
\frac{g(-S_{k},z_{k})}{\psi_{k}}\leq\dfrac{\mathcal{H}_{k}}{\psi_{k}}%
-\dfrac{1}{48}\Lambda_{k}^{2}(6+S_{k})<0.
\]
Since $g(0,z_{k})/\psi_{k}=\mathcal{H}_{k}/\psi_{k}>0,$ the Intermediate Value
Theorem implies that there exists a solution $s_{k}^{-}\in(-S_{k},0).$ The
monotonicity of $g(s,z_{k})$ by Lemma \ref{monotonicity}(ii) implies the
solution is unique in $(-S_{k},0)$ and claim (iv) is established.

Now assume $S_{k}<\frac{6}{5}$ and $0<\mathcal{H}_{k}/\psi_{k}<\frac{1}%
{16}\Lambda_{k}^{2}(2-S_{k}).$ Inequality (\ref{g_bound_2b}) implies%
\[
\frac{g(S_{k},z_{k})}{\psi_{k}}\leq\dfrac{\mathcal{H}_{k}}{\psi_{k}}-\frac
{1}{16}\Lambda_{k}^{2}(2-S_{k})<0.
\]
Since $g(0,z_{k})/\psi_{k}=\mathcal{H}_{k}/\psi_{k}>0,$ there exists a
solution $s_{k}^{+}\ $which by the monotonicity of $g(s,z_{k})$ (Lemma
\ref{monotonicity}(ii)) is unique in $(0,S_{k})$ establishing claim (v). To
establish claim (vi), assume $0<\mathcal{H}_{k}/\psi_{k}<\frac{9}{125}\left(
\psi_{k}/\psi_{k}^{\prime}\right)  ^{2}$. If $S_{k}\geq\frac{6}{5},$ by
inequality (\ref{g_bound_2b})%
\[
\frac{g(\frac{6}{5},z_{k})}{\psi_{k}}\leq\dfrac{\mathcal{H}_{k}}{\psi_{k}%
}-\dfrac{9}{125}\left(  \dfrac{\psi_{k}}{\psi_{k}^{\prime}}\right)  ^{2}<0
\]
which implies there exists a solution $s_{k}^{+},$ which by the monotonicity
of $g(s,z_{k}),$ is unique in $(0,\frac{6}{5})$ establishing claim (vi)a.
Moreover, if $S_{k}\geq6,$ inequality (\ref{g_bound_2b}) implies%
\[
0<\dfrac{\mathcal{H}_{k}}{\psi_{k}}\leq\dfrac{\mathcal{H}_{k}}{\psi_{k}%
}-\dfrac{1}{48}\Lambda_{k}^{2}(6-S_{k})\leq\frac{g(S_{k},z_{k})}{\psi_{k}}%
\]
which implies there must exist another solution $s_{k}^{\ast}\in(\frac{6}%
{5},S_{k})$ establishing claim (vi)b. Finally, the minimum value of
$\mathcal{H}_{k}/\psi_{k}-\frac{1}{48}\lambda^{2}(6-s)$ for $s>0$ is
$\mathcal{H}_{k}/\psi_{k}-\frac{2}{3}\left(  \psi_{k}/\psi_{k}^{\prime
}\right)  ^{2}.$ If $\mathcal{H}_{k}/\psi_{k}>\frac{2}{3}\left(  \psi_{k}%
/\psi_{k}^{\prime}\right)  ^{2},$ then using inequality (\ref{g_bound_2b}) we
have that for all $0<s<S_{k},$%
\[
0<\dfrac{\mathcal{H}_{k}}{\psi_{k}}-\dfrac{2}{3}\left(  \dfrac{\psi_{k}}%
{\psi_{k}^{\prime}}\right)  ^{2}\leq\frac{\mathcal{H}_{k}}{\psi_{k}}-\frac
{1}{48}\lambda^{2}(6-s)\leq\frac{g(s,z_{k})}{\psi_{k}}.
\]
Therefore, no solution can exist on $(0,S_{k})$, establishing claim (vii).
\end{proof}

\subsection*{Theorem \ref{EU_3}}

\begin{proof}
Since $\left\vert \lambda\right\vert \leq\Lambda_{k}<\left\vert \psi
_{k}^{\prime}\right\vert /48K,$ we have from equation (\ref{g_bound}) of Lemma
\ref{g_taylor} that
\[
\left\vert \frac{g(\lambda,z_{k})}{\psi_{k}^{\prime}}-\left(  \dfrac
{\mathcal{H}_{k}}{\psi_{k}^{\prime}}-\frac{1}{24}\lambda^{3}\right)
\right\vert \leq\frac{K}{\left\vert \psi_{k}^{\prime}\right\vert }\left\vert
\lambda\right\vert ^{4}<\frac{1}{48}\left\vert \lambda\right\vert ^{3}.
\]
If $\lambda\leq0,$ we have%
\begin{equation}
\dfrac{\mathcal{H}_{k}}{\psi_{k}^{\prime}}-\frac{1}{48}\lambda^{3}\leq
\frac{g(\lambda,z_{k})}{\psi_{k}^{\prime}}\leq\dfrac{\mathcal{H}_{k}}{\psi
_{k}^{\prime}}-\frac{1}{16}\lambda^{3}. \label{g_bound_3b}%
\end{equation}
If $\lambda\geq0,$ we have%
\begin{equation}
\dfrac{\mathcal{H}_{k}}{\psi_{k}^{\prime}}-\frac{1}{16}\lambda^{3}\leq
\frac{g(\lambda,z_{k})}{\psi_{k}^{\prime}}\leq\dfrac{\mathcal{H}_{k}}{\psi
_{k}^{\prime}}-\frac{1}{48}\lambda^{3}. \label{g_bound_3a}%
\end{equation}
To establish claim (i), assume $-\tfrac{1}{48}\Lambda_{k}^{3}<\mathcal{H}%
_{k}/\psi_{k}^{\prime}<0.$ Then if $\lambda<0,$ inequality (\ref{g_bound_3b})
implies
\[
0<\dfrac{\mathcal{H}_{k}}{\psi_{k}^{\prime}}+\dfrac{1}{48}\Lambda_{k}^{3}%
\leq\frac{g(-\Lambda_{k},z_{k})}{\psi_{k}^{\prime}}.
\]
Since $g(0,z_{k})/\psi_{k}^{\prime}=\mathcal{H}_{k}/\psi_{k}^{\prime}<0,$ the
Intermediate Value Theorem implies there must exist a solution $\lambda
_{k}^{-}\in(-\Lambda_{k},0).$ Uniqueness follows from monotonicity (Lemma
\ref{monotonicity}(iii)). Inequality (\ref{g_bound_3a}) implies that for all
$0<\lambda<\Lambda_{k},$
\[
\frac{g(\lambda,z_{k})}{\psi_{k}^{\prime}}\leq-\left\vert \dfrac
{\mathcal{H}_{k}}{\psi_{k}^{\prime}}\right\vert -\frac{1}{48}\left\vert
\lambda\right\vert ^{3}<0
\]
and thus no solution can exist on $(0,\Lambda_{k})$ establishing claim (i). A
parallel argument establishes claim (ii). The lower bound of inequality
(\ref{g_bound_3b}) and the upper bound of inequality (\ref{g_bound_3a})
establishes claim (iii). Finally, to establish claim (iv), assume
$\mathcal{H}_{k}/\psi_{k}^{\prime}<-\tfrac{1}{16}\Lambda_{k}^{3}.$ If
$-\Lambda_{k}\leq\lambda\,\leq0,$ inequality (\ref{g_bound_3b}) implies%
\[
\frac{g(\lambda,z_{k})}{\psi_{k}^{\prime}}\leq\dfrac{\mathcal{H}_{k}}{\psi
_{k}^{\prime}}+\frac{1}{16}\left\vert \lambda\right\vert ^{3}<\dfrac
{\mathcal{H}_{k}}{\psi_{k}^{\prime}}+\frac{1}{16}\Lambda_{k}^{3}<0
\]
and no solution can exist on $[-\Lambda_{k},0].$ Since by assumption,
$\mathcal{H}_{k}/\psi_{k}^{\prime}<0,$ the upper bound of inequality
(\ref{g_bound_3a}) implies no solution can exist on $[0,\Lambda_{k}].$ If, on
the other hand, $\mathcal{H}_{k}/\psi_{k}^{\prime}>\tfrac{1}{16}\Lambda
_{k}^{3},$ a parallel argument also implies no solution can exist and thus
claim (iv) is established.
\end{proof}

\bibliographystyle{hplain}
\bibliography{DTH_dynamics}

\begin{thebibliography}{10}

\bibitem{Chen-03}
Jing-Bo Chen, Han-Ying Guo, and Ke~Wu.
\newblock Total variation in {H}amiltonian formalism and symplectic-energy
  integrators.
\newblock {\em Journal of Mathematical Physics}, 44, April 2003,
  arXiv:hep-th/0111185.

\bibitem{Desbrun-05}
Mathieu Desbrun, Anil~N. Hirani, Melvin Leok, and Jerrold~E. Marsden.
\newblock Discrete exterior calculus.
\newblock August 2005, arXiv:math.DG/0508341 v2.

\bibitem{DInnocenzo-87}
A.~D'Innocenzo, L.~Renna, and P.~Rotelli.
\newblock Some studies in discrete mechanics.
\newblock {\em European Journal of Physics}, 8:245--252, 1987.

\bibitem{Ge-91}
Zhong Ge.
\newblock Equivariant symplectic difference schemes and generating functions.
\newblock {\em Physica D}, 49:376--386, 1991.

\bibitem{Ge-88}
Zhong Ge and Jerrold~E. Marsden.
\newblock Lie-{P}oisson integrators and {L}ie-{P}oisson {H}amiltonian-{J}acobi
  theory.
\newblock {\em Physics Letters A}, 133:134--139, 1988.

\bibitem{Goldstein-80}
Herbert Goldstein.
\newblock {\em Classical Mechanics}.
\newblock Addison-Wesley, 1980.

\bibitem{Golub-89}
Gene~H. Golub and Charles F.~Van Loan.
\newblock {\em Matrix Computations}.
\newblock Johns Hopkins University Press, 1989.

\bibitem{Greenspan-74}
Donald Greenspan.
\newblock {\em Discrete Numerical Methods in Physics and Engineering}.
\newblock Academic Press, 1974.

\bibitem{Greenspan-80}
Donald Greenspan.
\newblock {\em Arithmetic Applied Mathematics}.
\newblock Pergamon Press, 1980.

\bibitem{Greenspan-84}
Donald Greenspan.
\newblock Conservative numerical methods for $x''=f(x)$.
\newblock {\em Journal of Compuational Physics}, 56(1):28--41, 1984.

\bibitem{Guibout-04}
V.M. Guibout and A.~Bloch.
\newblock Discrete variational principles and {H}amilton-{J}acobi theory for
  mechanical systems and optimal control problems.
\newblock September 2004, arXiv:math.DS/0409296.

\bibitem{Hairer-97}
Ernst Hairer.
\newblock Variable time step integration with symplectic methods.
\newblock {\em Applied Numerical Mathematics}, 25:219--227, 1997.

\bibitem{Kane-99}
C.~Kane, J.E. Marsden, and M.~Ortiz.
\newblock Symplectic-energy-momentum preserving variational integrators.
\newblock {\em Journal of Mathematical Physics}, 40, July 1999.

\bibitem{Lanczos-70}
Cornelius Lanczos.
\newblock {\em The Variational Principles of Mechanics}.
\newblock Dover Publications, 1970.

\bibitem{Lee-83}
T.~D. Lee.
\newblock Can time be a discrete dynamic variable?
\newblock {\em Physics Letters Physics}, 122B:217--220, 1983.

\bibitem{Lee-87}
T.~D. Lee.
\newblock Difference equations and conservation laws.
\newblock {\em Journal of Statistical Physics}, 46:843--860, 1987.

\bibitem{Ortega-72}
James~M. Ortega.
\newblock {\em Numerical Analysis A Second Course}.
\newblock Academic Press, 1972.

\bibitem{Shibberu-92}
Yosi Shibberu.
\newblock {\em Discrete-Time {H}amiltonian Dynamics}.
\newblock PhD thesis, Univ. of Texas at Arlington, 1992,
  www.rose-hulman.edu/\linebreak[0]$\sim$shibberu/\linebreak[0]DTH\_Dynamics/D%
TH\_Dynamics.htm.

\bibitem{Shibberu-94}
Yosi Shibberu.
\newblock Time-discretization of {H}amiltonian dynamical systems.
\newblock {\em Computers and Mathematics with Applications},
  28(10-12):123--145, 1994.

\bibitem{Shibberu-97}
Yosi Shibberu.
\newblock A discrete-time formulation of {H}amiltonian dynamics.
\newblock June 1997,
  www.rose-hulman.edu/\linebreak[0]$\sim$shibberu/\linebreak[0]DTH\_Dynamics/D%
TH\_Dynamics.htm.

\bibitem{Shibberu-98}
Yosi Shibberu.
\newblock A discrete-time formulation of {H}amiltonian dynamics.
\newblock February 1998,
  www.rose-hulman.edu/\linebreak[0]$\sim$shibberu/\linebreak[0]DTH\_Dynamics/D%
TH\_Dynamics.htm.

\bibitem{Shibberu-05}
Yosi Shibberu.
\newblock How to regularize a symplectic-energy-momentum integrator.
\newblock July 2005,
  www.rose-hulman.edu/\linebreak[0]$\sim$shibberu/\linebreak[0]DTH\_Dynamics/D%
TH\_Dynamics.htm.

\bibitem{Talasila-04}
V.~Talasila, J.~Clemente-Gallardo, and A.~J. van~der Schaft.
\newblock Geometry and {H}amiltonian mechanics on discrete spaces.
\newblock {\em Journal of Physics A: Mathematical and General}, 37:9705--9734,
  2004.

\end{thebibliography}

\end{document}